\documentclass[twocolumn,pra,aps,amsmath,amssymb,superscriptaddress,showpacs]{revtex4-1}

\usepackage{amsfonts}
\usepackage{amsmath}
\usepackage{amssymb}
\usepackage{mathbbol}
\usepackage{graphicx}
\usepackage{times}
\usepackage[english, activeacute]{babel} 
\usepackage[utf8]{inputenc} 
\usepackage{color}
\newcommand{\new}[1]{{\color{black}{#1}}}

\newcommand{\comment}[1]{\textbf{\color{blue}{[#1]}}}
\newcommand{\insertrefs}{\comment{Refs.}}

\newcommand{\old}[1]{}

\begin{document}

\title{Time-Dependent Density-Functional Theory of Strong-Field Ionization of Atoms under Soft X-Rays}

\author{A. Crawford-Uranga}
\email{alisonc1986@gmail.com}
\affiliation{Nano-Bio Spectroscopy Group and ETSF Scientific Development Center, Departamento de F\'isica de Materiales, Centro de F\'isica de Materiales CSIC-MPC and DIPC, Universidad del Pa\'is Vasco UPV/EHU, Avenida de Tolosa 72, E-20018, San Sebasti\'an, Spain}
\author{U. De Giovannini}
\affiliation{Nano-Bio Spectroscopy Group and ETSF Scientific Development Center, Departamento de F\'isica de Materiales, Centro de F\'isica de Materiales CSIC-MPC and DIPC, Universidad del Pa\'is Vasco UPV/EHU, Avenida de Tolosa 72, E-20018, San Sebasti\'an, Spain}
\author{E. R\"{a}s\"{a}nen}
\affiliation{Department of Physics, Tampere University of Technology, FI-33101 Tampere, Finland}
\author{M. J. T. Oliveira}
\affiliation{Center for Computational Physics, University of Coimbra, Rua Larga, 3004-516 Coimbra, Portugal}
\author{D. J. Mowbray}
\affiliation{Nano-Bio Spectroscopy Group and ETSF Scientific Development Center, Departamento de F\'isica de Materiales, Centro de F\'isica de Materiales CSIC-MPC and DIPC, Universidad del Pa\'is Vasco UPV/EHU, Avenida de Tolosa 72, E-20018, San Sebasti\'an, Spain}
\author{G.~M.~Nikolopoulos}
\affiliation{Institute of Electronic Structure and Laser, FORTH, P.O. Box 1527, GR-71110 Heraklion, Greece}
\author{E. \new{T.} Karamatskos}
\thanks{Present address: Institut f\"ur Laser-Physik, Universit\"at Hamburg, Luruper Chaussee 149, D-22761 Hamburg, Germany.}
\affiliation{Department of Physics, University of Crete, P.O. Box 2208, GR-71003 Heraklion, Crete, Greece}
\author{D. Markellos}
\affiliation{Department of Physics, University of Crete, P.O. Box 2208, GR-71003 Heraklion, Crete, Greece}
\author{P. Lambropoulos}
\affiliation{Institute of Electronic Structure and Laser, FORTH, P.O. Box 1527, GR-71110 Heraklion, Greece}
\affiliation{Department of Physics, University of Crete, P.O. Box 2208, GR-71003 Heraklion, Crete, Greece}
\author{S. Kurth}
\affiliation{Nano-Bio Spectroscopy Group and ETSF Scientific Development Center, Departamento de F\'isica de Materiales, Centro de F\'isica de Materiales CSIC-MPC and DIPC, Universidad del Pa\'is Vasco UPV/EHU, Avenida de Tolosa 72, E-20018, San Sebasti\'an, Spain}
\affiliation{IKERBASQUE, Basque Foundation for Science, E-48011, Bilbao, Spain}
\author{A. Rubio}
\email{angel.rubio@ehu.es}
\affiliation{Nano-Bio Spectroscopy Group and ETSF Scientific Development Center, Departamento de F\'isica de Materiales, Centro de F\'isica de Materiales CSIC-MPC and DIPC, Universidad del Pa\'is Vasco UPV/EHU, Avenida de Tolosa 72, E-20018, San Sebasti\'an, Spain}

\date{\today}

\begin{abstract}
We demonstrate the capabilities of time-dependent density functional theory
(TDDFT) for strong-field, short wavelength (soft X-ray) physics, as compared to 
a formalism based on rate equations. 
We \new{find} that TDDFT provides a very good description 
of the total and individual ionization yields for Ne and Ar atoms
exposed to strong laser pulses.  
We assess the reliability of different adiabatic density functionals and conclude that \new{an accurate description of long-range interactions by} 
the exchange and correlation \new{potential}
\new{is} crucial for obtaining the correct ionization yield over
a wide range of intensities ($10^{13}$ -- $\new{5\times}10^{15}$ W/cm$^2$). 
Our TDDFT calculations disentangle the contribution from each ionization channel \new{based on}
the Kohn-Sham wavefunction\new{s}.
\end{abstract}


\maketitle 

\section{Introduction}\label{intro:sec}

Recent advances in the development of free electron lasers (FELs) have 
led to the generation of intense, \new{ultra}short duration, and short wavelength radiation sources 
ranging from extreme ultraviolet (XUV) 
to hard X-rays~\cite{go1,1}. 
\new{The possible applications encompass a broad area of
topics, such as basic atomic and molecular physics, dense matter, and imaging of complex
biomolecules, to mention only a few~\cite{AND_MORE, LAM_AGAIN_AGAIN}. 
As radiation in this wavelength range can ionize deep inner electrons, 
multiple ionization processes are an inevitable outcome. Understanding the
mechanism underlying these processes~\cite{2,21,3,4,41,LAMBRO,LAMBRO1} is of fundamental importance to this broad
interdisciplinary field.}

\new{For infrared to optical high intensity lasers, even for ultra-short pulse durations
down to a few cycles, the validity of the single active electron (SAE) approximation
is well established~\cite{LAM_AGAIN}. As a consequence, multiple ionization is dominated by
sequential stripping of valence electrons. To the extent that, for sufficiently high
peak intensities, with ponderomotive energy much larger than the photon energy
and Keldysh parameter $\gamma$$\ll$1, non-sequential two or three electron escape may be
observable, the recollision mechanism has been shown to provide a valid
description~\cite{LAM_AGAIN}. However, even for intense long wavelength radiation, it is still
difficult to produce highly charged ions.}

\old{For infrared to optical high intensity lasers, even for
 ultrashort pulse durations down to a few field cycles ~\cite{10,101,11}, 
the validity of the single active electron (SAE) approximation is well established~\insertrefs. 
Here, multiple ionization is dominated by
sequential stripping of electrons~\cite{7,71,8,81,9}. 
For sufficiently high laser intensities, where the ponderomotive
energy is much larger than the photon energy, and the Keldysh parameter $\gamma$$\ll$1, non-sequential two or
three electron escape processes are also possible~\insertrefs. 
In this case, the recollision mechanism has been shown to provide a valid description.
However, even for intense long wavelength radiation it is still difficult to produce highly charged ions \insertrefs.}

On the other hand, short wavelength FEL radiation, more often than not, produces highly charged ions in abundance~\cite{2,21,3,4,41, NATURE_PHOT}. Typically, the stripping begins with the ejection of one or more subvalence electrons, but the physical processes that determine the course of events depend strongly on the photon energy range. For hard X-rays, say above 2 keV, it is mainly single-photon inner shell electron ejections, followed by avalanches of Auger decays and rearrangement, that dominate. By the time  highly ionized species appear, with ionization potentials higher than the photon energies, the pulse is essentially over, which minimizes the possibility of two- or multi-photon ionization~\cite{NATURE_PHOT}. However, for soft X-ray energies, say up to 300 eV or so, the single-photon subvalence  ionization  eventually mingles, with (non-linear) multiphoton processes providing thus an unusual and theoretically demanding interplay between linear and non-linear processes. For currently accessible FEL peak intensities, for which the ponderomotive energy is much smaller than the photon energy, consistent with $\gamma$$\gg$1, and pulse durations of hundreds of field cycles, lowest non-vanishing order perturbation theory (LOPT), in terms of rate equations and multiphoton cross sections is a valid model~\cite{4}, in the entire FEL photon energy range.

Although sequential ionization still plays a dominant role, an entirely different  non-sequential mechanism of multiple ionization comes into play. The SAE and recollision-based models are totally inapplicable in this context, because sequential ionization begins with subvalence electrons, for which the relevant cross sections, be it single- or multi-photon, involve inter- as well as intra-shell correlation. Additionally, even within LOPT, the calculation of multiphoton cross sections, which requires explicit or implicit summations over complete sets of intermediate states, poses a formidable computational challenge; not to mention the further complexity introduced by the possibility of multiphoton multielectron escape~\cite{41,LAMBRO}. Experimentation with alternative methods, circumventing this task, is therefore highly desirable. 

As FEL intensities are expected to increase and pulse durations be shortened, LOPT is expected to lose its validity. In the soft X-ray regime, this is apt to occur for peak intensities above $10^{17}$ W/cm$^2$ and pulse durations well below 5 fs at which point a non-perturbative approach will become necessary. As solving the time dependent Schrodinger equation beyond the SAE approximation is a daunting task even for two electrons~\cite{DAUNTING, DAUNTING_1}, time dependent density functional theory (TDDFT) appears to be one of the few available options. Since this represents uncharted territory, we have chosen to assess the potential of the method by applying TDDFT to the calculation of total ionization, as well as individual ionic yields for Ne and Ar, under photon energies 93 and 105 eV, respectively. In both cases, we have chosen peak intensities for which LOPT is demonstrably valid, in addition to the availability of some experimental data~\cite{LAMBRO1,NEW_REF}. We are thus in the position to obtain a first assessment of the potential of TDDFT, before venturing into the non-perturbative regime. 

Applications of TDDFT in the long-wavelength strong-field regime are relatively scarce. Although steps in that direction, with mixed success, were already taken in the 1990s~\cite{STRONG, STRONG1, STRONG2, Lappas:1998gc, Lein:2000cu}, to the best of our knowledge, this is the first time TDDFT has been used in the interaction  of atoms with short wavelength FEL radiation. This is probably because TDDFT was previously thought to fail to describe strong field ionization under IR radiation, as it had limited success in accounting for the so-called knee in helium double ionization~\cite{KNEE, KNEE_AGAIN, LAM_AGAIN} . Instead, as discussed in the sections that follow, we find that TDDFT actually does provide a surprisingly good description of several aspects of the non-linear dynamics of atoms driven by strong soft X-ray radiation. The remaining discrepancies between LOPT and TDDFT provide a road map towards further improvement, in preparation for the extension of the approach to shorter wavelengths and /or more complex systems.

In Sec. \ref{theory:sec} we briefly present the theoretical methodology to study the ionisation of Ne and Ar. We first introduce in Sec. \ref{TDDFT_LOPT:sec} the theoretical approach that we have used to calculate the ionisation yields. 
We then show in Sec. \ref{laser:sec} how we model the laser field that we apply to our atoms. In Sec. \ref{yields:sec} we show how we obtain both the total and individual yields.  Finally, in Sec. \ref{numerical_details:sec}, we provide numerical details of how we perform the calculations, including Appendix \ref{AppendixA}, to which the reader can refer to for more details.

In Sec. \ref{results_discussion:sec} we present the results and discussion 
for the total and individual yields obtained for Ar and Ne atoms exposed to a 
strong-field, short wavelength (soft X-ray) laser, as a function of the laser intensity. This is followed by concluding remarks in Sec. \ref{conclusions:sec}.

\section{Theoretical Background}\label{theory:sec}

\subsection{Time-Dependent Density-Functional Theory}\label{TDDFT_LOPT:sec}

The central tenet of TDDFT is that all physical 
properties of an interacting many-electron system can be 
determined from its time-dependent density~\cite{RUNGEGROSS}. As in static 
DFT~\cite{KS}, the interacting system is mapped -- in principle exactly -- 
\new{o}nto an auxiliary, non-interacting system, the so-called Kohn-Sham (KS) 
system, which by construction yields the same time-dependent density as the 
interacting one. 
In the present work we are interested in the non-linear dynamics of closed-shell electronic 
systems, and 
for $N_0$ electrons the density of the KS system is 
$n(\mathbf{r}, t) = 2 \sum_{i=1}^{N_0/2} {|{\phi_i (\mathbf{r}, t)|}}^2$
(we here assume a spinless ground state),
where $\phi_i (\mathbf{r}, t)$ are single-particle KS orbitals satisfying  
the time-dependent KS (TDKS) equations (in atomic units):
\begin{eqnarray}
i \frac{\partial}{\partial t} \phi_i (\mathbf{r}, t) =&& \bigg[ 
-\frac{\nabla^2}{2} + V_{0}(\mathbf{r}) 
+ V_{\rm FEL}(\mathbf{r}, t) + V_{\rm H}[n](\mathbf{r}, t) 
\bigg. \nonumber \\
&&  + V_{\mathit{xc}}[n](\mathbf{r}, t) \bigg] 
\phi_i (\mathbf{r}, t),\,\,\,  i=1,\dots,N_0/2.
\label{tdks:eqn}
\end{eqnarray} 
Here $V_0(\mathbf{r})$ is the electrostatic potential of the 
nuclei, $V_{\rm FEL}(\mathbf{r}, t)$ describes the laser field, 
$V_{\rm H}[n](\mathbf{r}, t)$ 
is the Hartree potential and $V_{\mathit{xc}}[n](\mathbf{r}, t)$ is the exchange-correlation (xc) 
potential.
In this work we consider different xc~potential approximations: 
the local-density approximation~\cite{LDA1} (LDA), PBE~\cite{PBE} 
and LB94~\cite{LB94} forms of the generalized gradient approximations,
and the corrected-exchange-density~\cite{XCLDA} extension of LDA (CXD-LDA).
Further, we employ the adiabatic extension of these functionals to the time dependent case.

\subsection{Laser field}\label{laser:sec}

We model the laser-atom interaction within the dipole approximation 
using an external potential defined as 
\begin{equation}\label{eq:vfel}
V_{\rm FEL}(t) = A f(t) \sin (\omega t) \mathbf{r}\cdot \mbox{\boldmath$\alpha$} 
\end{equation}
where $\mbox{\boldmath$\alpha$}$ is the polarization, $\omega$ the frequency 
and $A$ the field amplitude of the laser. 
This approximation is well justified under FEL experimental conditions.
The pulse envelope is of Gaussian shape $f(t) = \exp\left[-(t-t_0)^2/2\tau_0^2\right]$ , 
with peak value centered at $t_0$ and full width at half maximum (FWHM) intensity 
given by $2\sqrt{\ln2}\tau_0$.
A deterministic temporal shape as defined in Eq.~(\ref{eq:vfel})
represents a simplified model compared to current FEL sources 
which, in general, exhibit strong intensity fluctuations.

\subsection{Total and individual yields}\label{yields:sec}

In order to estimate the total ionic yields, we follow the time 
evolution of the charge remaining in a given volume $V$ around the atom. 
The norm of each KS orbital inside this volume $N_{i}(t)=\int_{V} {\rm d} 
\mathbf{r} \, |\phi_i(\mathbf{r},t)|^2$ decreases in time during the 
application of a strong short wavelength laser pulse. 
The total number of escaped electrons at time $t$ is given by
\begin{equation}
N_{\rm esc}(t) = N_0 - N(t)\,.
\label{TDDFT_eqn}
\end{equation}
where $N_0$ is the initial number of electrons and $N(t)=\sum_{i}N_{i}(t)= \int_{V} {\rm d} \mathbf{r} \, n(\mathbf{r},t)$, 
is the total number of remaining electrons in the volume at the given time.
The total ionization yield is the long time limit of Eq.~(\ref{TDDFT_eqn}): 
$N_{\rm esc}=N_{\rm esc}(t\rightarrow\infty)$.

To calculate the individual ionization probabilities 
$P^{q+}(t)$ for an atom, i.e., the probability to produce an ion in a positively $q$-charged 
state ($q=1,\dots, N_0$), we employ the approach based on the time-dependent 
KS orbitals described in Ref.~\cite{Ullrich:2000tm}.
In this approximation, the ionization probability is defined as the 
sum over all the combination\new{s} of KS probabilities $N_i(t)$ composing a $q$-charged state 
\begin{equation}
P^{q+}(t)= \sum_{\substack{\sigma\in \mathbf{C}(N_0,q) }} N_{\sigma(1)}(t) \cdots
N_{\sigma(N-q)}(t),  
\label{ion_ch}
\end{equation}
where $\mathbf{C}(N_0,q)$ is the set of all the possible configurations $\sigma$ in which an 
($N_0-q$)-tuple can be selected from a ($1,\dots,N_0$) one. 
The total ionization yield can be reconstructed by computing a weighted sum over 
the different ionization channels.
From LOPT we directly obtain all the individual ionic yields $P^{q+}$~\cite{LAMBRO, LAMBRO1} 
and the total yield can be obtained with the same prescription.
Within TDDFT, we can directly obtain the total yield either
by summing up the individual ionization channels or by monitoring the total charge.  
In either case, the total yield is the only quantity that is rigorously correct since 
it can be directly derived from the total density.
To obtain the individual ionic channels, we need to assume that the KS wavefunction 
is a good representation of the exact many-body one. The validity of this assumption in 
the present context has to be evaluated on the basis of its success in recapturing results from
different approaches and, of course, experimental data.  

\subsection{Numerical details}\label{numerical_details:sec}

In the following we present results for strong field ionization of \new{Ne} and \new{Ar} atoms.
We numerically propagate the TDKS equations~(\ref{tdks:eqn}) in real time and real space
using the {\sc octopus} code~\cite{OCTOPUS1, OCTOPUS2, OCTOPUS3}. 
To this end we perform calculations in a 16~${\rm\AA}$ radius spherical box 
and discretize the problem on a cartesian grid with spacing 0.16~${\rm\AA}$. 
An 8~${\rm\AA}$ thick boundary absorber~\cite{PUBLISH} is introduced to account for
electrons escaping from the simulation box, which therefore acts as an integration volume $V$.
Core excitations are expected to play a relevant role for intensities much higher 
than the ones considered in this paper~\cite{LAMBRO, LAMBRO1}. 
For this reason we freeze the core electrons into a pseudopotential using the Troullier-Martins
scheme~\cite{TM} for both atoms\new{:} 1s electrons for Ne and 1s, 2s, \new{and} 2p \new{electrons} for Ar \cite{APE}.
We refer to Appendix \ref{AppendixA} for more information regarding an assessment of 
the pseudopotential\new{, the grid parameters} and the absorbing boundary used here.    

\begin{figure}
\includegraphics[width=\columnwidth]{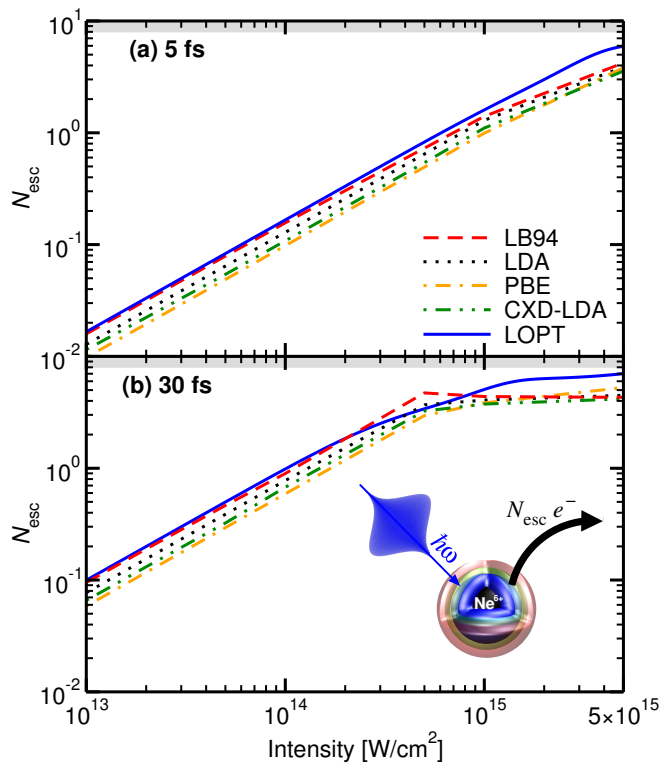}
\caption{
(color online) \new{Ne} total number of escaped electrons $N_{\mathrm{esc}}$ for different 
laser intensities and (a) 5 fs and (b) 30 fs FWHM pulses of $\omega=93$~eV.
Different TDDFT functionals are compared with LOPT. 
\new{The \new{Ne} ionization process is shown as an inset.}
Shaded regions indicate the electrons frozen in the pseudopotential.
}
\label{NE_5fs:fig}
\end{figure}  

\section{Results and discussion}\label{results_discussion:sec}

\subsection{Ne atom}\label{neon:sec}

In Fig.~\ref{NE_5fs:fig} we compare TDDFT and LOPT 
total ionization yields for \new{Ne} as a function of the laser intensity\new{, as depicted schematically in the inset}.
Here, the photon energy is fixed to $\omega=93$~eV and we consider
the cases of a short 5~fs FWHM - (a), and a 30~fs FWHM long laser 
pulse - (b). 
In the two cases we propagate in time the TDDFT equations 
for 25~fs and 153~fs respectively\new{. We then use an exponential
fit near the end of the propagation to extrapolate the total ionization yield.}
The overall agreement is remarkably good for all the xc~functionals
in a wide range of intensities (plots are in log scale).
However, as the intensity increases, 
the agreement gradually deteriorates with TDDFT tending towards lower ionization yields.
This behavior is more pronounced in the  
30~fs case in Fig.~\ref{NE_5fs:fig}~(b), where the TDDFT ion yield flattens out for intensities
$\gtrsim 10^{15}$ W/cm$^2$, while LOPT yields higher values. The observed deviation \new{may trace} back
to an improper time-dependence of the xc \new{potential} for highly ionized species\new{, 
as has been proven in 1D for He, a two electron system~\cite{Fuks:2011hp, Fuks:2013fn, PRL_ELLIOT}}. 
These effects become more severe for high intensities and build up
for longer times because more electrons are ejected and the density changes are more
substantial. For systems with more electrons we expect them to introduce some additional
dynamical screening that can change the magnitude of these effects but not their presence.
Consequently, there are spurious oscillations in the successive KS ionization potentials 
leading to increasing errors as the system loses electrons (see below).
An error $\lesssim 10$\% in the KS ionization potentials brought in by 
the pseudopotential for the strongly charged ions cannot alone justify the observed 
effect.   

As LOPT has shown itself to be in excellent agreement with experiment~\cite{4,41,LAMBRO,LAMBRO1},
we deduce that our TDDFT results have a tendency to slightly underestimate the total ionic yield. 
It must be added, however, that the\new{se} difference\new{s are}  minor and \new{are} likely to 
fall within the \new{present} experimental accuracy of many FEL experiments. 
Therefore\new{,} we  conclude that TDDFT has predictive power \new{over} a wide range of laser pulse intensities.

Not all the xc~functionals perform in the same way.
A characterizing property of both LB94 and CXD-LDA is the correct asymptotic 
tail decay $V_{\mathit{xc}} \sim -1/r$ following \new{the }Coulomb \new{potential} for large $r$.
In contrast, both LDA and PBE decay exponentially.
The high-lying unoccupied KS bound states, close to the ionization threshold, 
are thus expected to be more accurately described by the LB94 and CXD-LDA functionals.
\new{This} is reflected \new{in} a superior description of the ionization process with LB94, as it provides the best \new{agreement with} LOPT. 
In this respect, the relatively poor accuracy of CXD-LDA compared \new{to} LB94
deserves further examination.

\begin{figure}
\includegraphics[width=\columnwidth]{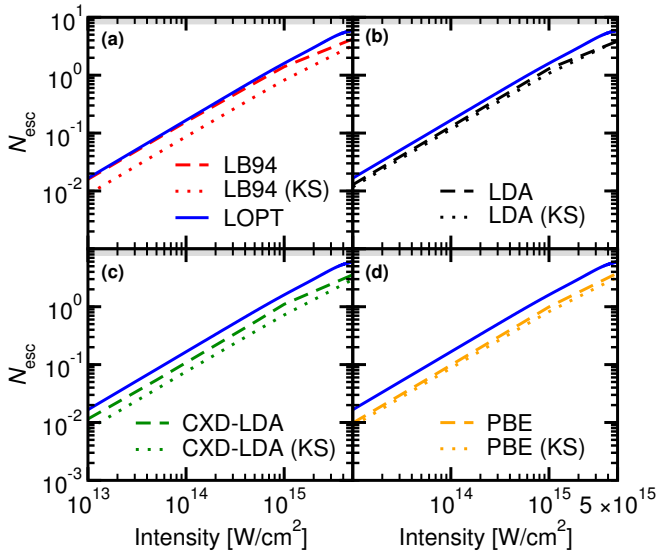}
\caption{
(color online) As in Fig.~\ref{NE_5fs:fig} for a 5~fs FWHM pulse
and different approximation levels:
LB94 (a), LDA (b), CXD-LDA (c) and PBE (d). 
In each panel we compare LOPT (solid), TDDFT (dashed) and 
\new{the independent KS response} \old{frozen (F)} (dotted).
Shaded regions indicate the electrons included in the pseudopotential.
}
\label{NE_FROZEN_VS_NON_FROZEN:fig}
\end{figure} 
To discern the impact of the underlying ground state and the 
quality of the Hartree plus exchange-correlation functional, 
we compare in Fig.~\ref{NE_FROZEN_VS_NON_FROZEN:fig} the 
full solution of the TDKS with the one in which we 
keep $V_{\rm H}$ and $V_{\mathit{xc}}$ frozen in the initial 
ground state configuration \new{(the independent KS response)}. 
Electrons are thus treated as non-interacting particles moving in a fixed 
external potential.
The effects of such a crude approximation are almost indiscernible 
\new{when the xc~potential is short ranged, i.e.\new{,} $V_{\mathit{xc}}  \sim e^{-r}$, as for} LDA and PBE (\emph{cf.}\ Fig.~\ref{NE_FROZEN_VS_NON_FROZEN:fig} (b) and (d)).
\new{However, this is not the case when the xc~potential is long-ranged, i.e.\new{,}  $V_{\mathit{xc}} \sim -1/r$, as for} LB94 and CXD-LDA (\emph{cf.}\ Fig.~\ref{NE_FROZEN_VS_NON_FROZEN:fig} (a) and (c)).
\new{

In an independent KS response picture, the total ionization yields are directly related to the KS eigenvalues.  For long-ranged xc~potentials, the KS eigenvalues are more strongly bound, reducing total ionizaton yields compared to short-ranged xc~potentials. When the xc~potential is propagated in time, ejected electrons may induce an attractive potential via redistribution of the electronic density on the ion.  Thus, the 
kinetic energy of the ejected electrons will be reduced due to this stabilization of the electronic levels. 
To support our analysis we have employed linear
response TDDFT to calculate the cross sections.
From Appendix \ref{AppendixA}, we see that the cross sections increase 
as the kinetic energies decrease below $93$~eV~\cite{CHU}, 
so we obtain a larger number of escaped electrons than when the xc~potential is frozen.
This effect becomes relevant mostly when the long-range electron-electron interaction is accurately described.}
This leads to a substantial modification in the number of electrons being ejected
($\sim$100\% increase for LB94 and $\sim$33\% for CXD-LDA at $I=10^{14}$~W/cm$^2$). 

\begin{figure}
\includegraphics[width=\columnwidth]{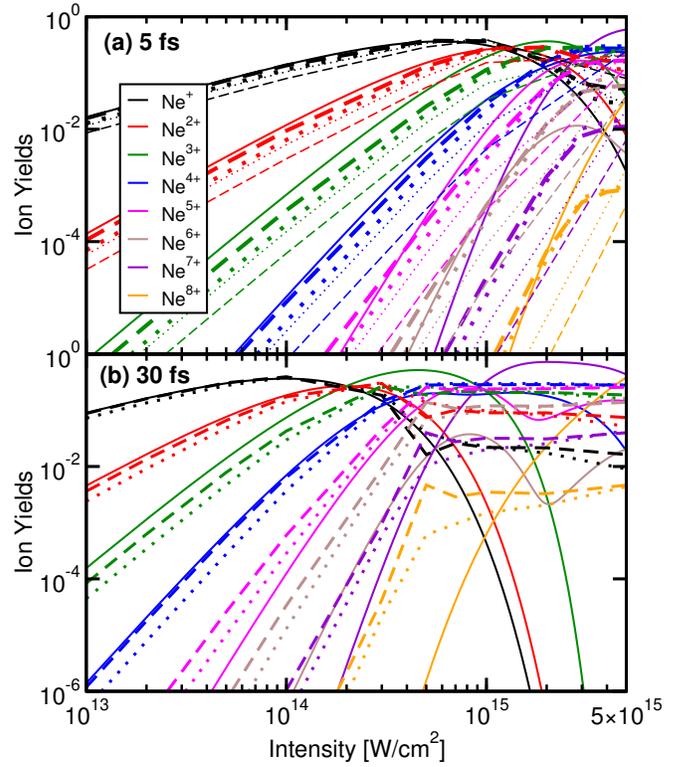}
\caption{
(color online) \new{Ne} individual ionization yields as a function of the intensity 
for (a) 5 fs and (b) 30 fs FWHM laser pulses of $\omega=93$~eV. 
TDDFT \new{(thicker) and the independent KS response (thinner) }with LB94 (dashed) and LDA (dotted) functionals are
compared to LOPT (solid). 
}
\label{NE_CHANNELS_5fs:fig}
\end{figure}
In Fig.~\ref{NE_CHANNELS_5fs:fig} we analyze \new{Ne} individual \new{TDDFT 
(thicker) and the independent KS response (thinner)} ionization 
yields obtained with LB94 and LDA using Eq.~(\ref{ion_ch}).
We observe that TDDFT ionic yields up to Ne$^{4+}$ are in good agreement with LOPT
for a large range of laser intensities; $I\lesssim 10^{15}$ W/cm$^2$ for 
a 5~fs pulse, [Fig.~\ref{NE_CHANNELS_5fs:fig}~(a)], and  $I\lesssim 5\times10^{14}$ W/cm$^2$ 
for a 30~fs one [Fig.~\ref{NE_CHANNELS_5fs:fig}~(b)].
For more strongly ionized species, Ne$^{5+}$ through Ne$^{8+}$, the discrepancy 
is larger, especially for the 30~fs pulse.
\new{From Fig.~\ref{NE_CHANNELS_5fs:fig}~(a) 
we see that the individual ionization yields for all channels 
are ordered as LB94(KS) $<$ LDA(KS) $\sim$ LDA $<$ LB94, 
as was also the case for the total ionization yields shown in Fig.~\ref{NE_FROZEN_VS_NON_FROZEN:fig} (a) and (b).}

The time-locality of the xc~\new{potential} introduces fluctuations that are amplified 
by the charge status of the ion \cite{PRL_ELLIOT}. 
As a consequence, the total and partial ionization yields are in good agreement with LOPT, 
as long as the channels with a charge status $\geq Ne^{5+}$ 
play a negligible role in the ionization process (\emph{cf.}\ Figs.~\ref{NE_5fs:fig} and  
\ref{NE_CHANNELS_5fs:fig}). 
Experimental ionization channels up to Ne$^{6+}$ present excellent 
agreement with LOPT~\cite{4,41,LAMBRO,LAMBRO1}.
We can therefore conclude that TDDFT describes well the ionic yields up to
Ne$^{4+}$, with the current state-of-the-art experimental data. 

The trend observed in the TDDFT total yields for the different 
xc~functionals, [Fig.~\ref{NE_5fs:fig}], 
is reflected in the single ionization channels for both pulse lengths:
namely, the inclusion of a correct asymptotic decay systematically 
improves the description of each channel.  

\begin{figure}
\includegraphics[width=\columnwidth]{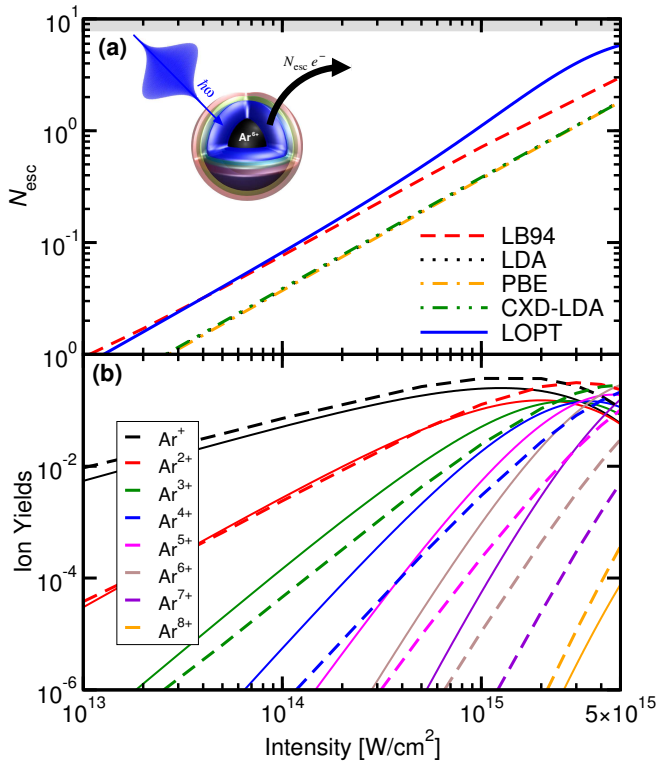}
\caption{
(color online) \new{Ar} total and individual ionic yields 
as a function of the laser intensity for a 10~fs FWHM laser pulse of $\omega=105$~eV.
(a) Total ionization yield for different TDDFT functionals and LOPT. 
(b) Individual ionization channels for LB94 (dashed) and LOPT (solid).
\new{The \new{Ar} ionization process is shown as an inset.}
Shaded regions indicate electrons frozen in the pseudopotential.
}
\label{AR_CHANNELS:fig}
\end{figure}  

\subsection{Ar atom}\label{argon:sec}

In Fig.~\ref{AR_CHANNELS:fig} we present the results for total and individual ionization yields
of \new{Ar}\new{, as depicted schematically in the inset,} for a 10 fs FWHM pulse of energy $\omega = 105$~eV and a full 
propagation time of 51~fs.
The total ionization yields for all the considered functionals except LB94,
[Fig.~\ref{AR_CHANNELS:fig}~(b)], qualitatively follow LOPT but systematically 
predict lower values. LB94 remarkably reproduces LOPT up to 
$I\lesssim 5\times10^{14}$ W/cm$^2$.
For higher intensities, it departs towards lower ionization values 
similarly to what was observed 
for \new{Ne} in Fig.~\ref{NE_5fs:fig}.

The intensity dependence of single ionization channels, as shown in
Fig~\ref{AR_CHANNELS:fig}~(b), is in good agreement up to  
Ar$^{3+}$ for LB94 only. It then quickly deteriorates for higher ionized species.

From experimental results~\cite{NEW_REF} we know that when the Ar$^{6+}$ is produced, 
it is composed of an ionized small contribution and a much
larger contribution on an excited state. However, its ionized
contribution is the predominant one to produce Ar$^{7+}$ and Ar$^{8+}$.
Additionally, the single ionization contribution to Ar$^{7+}$ is also much larger than the
double ionization contribution to Ar$^{8+}$.
The strongly ionized Ar$^{7+}$ is here produced through a sequential ionization process 
involving a shakeup step where a photon ejects one electron in the continuum 
while leaving the parent ion in an excited state~\cite{NEW_REF}.
Discarding this ionization pathway in LOPT leads to an Ar$^{7+}$ yield reduced up 
to \old{a factor of }\new{4 orders of} magnitude with respect to experiment.
In general, TDDFT tends to underestimate the LOPT results for highly charged channels. 
\new{The shake-up experimental effect for Ar$^{7+}$ is only partially accounted for in TDDFT, 
since none of the xc functionals we employed are self-interaction free.}
On the other hand, for Ne$^{8+}$ and Ar$^{8+}$, spurious
correlation effects between the core electrons frozen in the pseudopotential and the escaped
electrons in the absorbing boundary are included. This may lead to the overestimation of
TDDFT with respect to LOPT. For this reason, in TDDFT the Ar$^{7+}$ channel is strongly suppressed\new{, 
while the Ar$^{8+}$ channel is \old{strongly }enhanced}.
However, as we are most interested in the intensity regime
where LOPT is expected to be valid, these issues are not so relevant for this study.


\section{Conclusions}\label{conclusions:sec}

In conclusion, we have compared TDDFT and LOPT ionization yields for \new{Ar} and \new{Ne}, 
subjected to intense soft X-ray radiation.
Overall, for both short and long pulse durations, TDDFT results display \new{noteworthy}
similarities with LOPT throughout a wide range of intensities and 
within present-day experimental error.
Using a functional with the correct asymptotic behaviour significantly improves 
the potential applicability of TDDFT.
Since the two approaches are built on \new{completely} different bases, the resulting agreement 
indicate\new{s} that both are able to provide a realistic picture of the underlying physics. 
In particular, this results in the demonstration of an unexpected predictive power of TDDFT 
in describing total and individual ionization yields in FEL experiments. 
We believe that, in addition to the illustration of the predictive potential of TDDFT in the present context, 
this work introduces a road map for the exploration of non-perturbative approaches 
in short wavelength strong field physics, with a number of open questions to be addressed.


\section{Acknowledgements}\label{acknowledgements:sec}

We acknowledge financial support from the European Research Council Grant
DYNamo (ERC-2010-AdG-267374), Spanish Grant (FIS2010-21282-C02-01),
Grupos Consolidados UPV/EHU del Gobierno Vasco (IT-578-13), COST-CM1204 (XLIC),
and the European Commission project CRONOS (Grant number 280879-2). 
A.C.-U.\ acknowledges a fellowship from the Gobierno Vasco (Ref. BFI-2011-26); 
M.J.T.O.\ from the Portuguese FCT (contract No. SFRH/BPD/44608/2008);
E.R.\ from the Academy of Finland and 
D.J.M.\ from the Spanish Juan de la Cierva fellowship (JCI-2010-08156). 
We acknowledge the computational resources 
provided by the Red Española de Supercomputaci\'on.

\appendix

\section{Assessment of the pseudopotential\new{, the grid parameters} and the absorbing boundary }\label{AppendixA}

\subsection{Absorbing boundaries and continuum states description}\label{CONTINUUM:sec}

Modeling ionization processes involves a minute description of the 
interaction (through a laser pulse) between bound and continuum states.
A description of finite volume continuum states in real-space real-time propagation methods, 
is customarily achieved through the use of boundary absorbers \cite{PUBLISH}.
In the absence of such absorbers, the electronic wavepackets are reflected 
back and forth at the boundaries of the simulation box.  
Complex absorbing potentials (CAPs) constitute a widely used solution to 
eliminate such reflections~\cite{Leforestier:1983gka, Kosloff:1986eta}.
\begin{figure}
\includegraphics[width=\columnwidth]{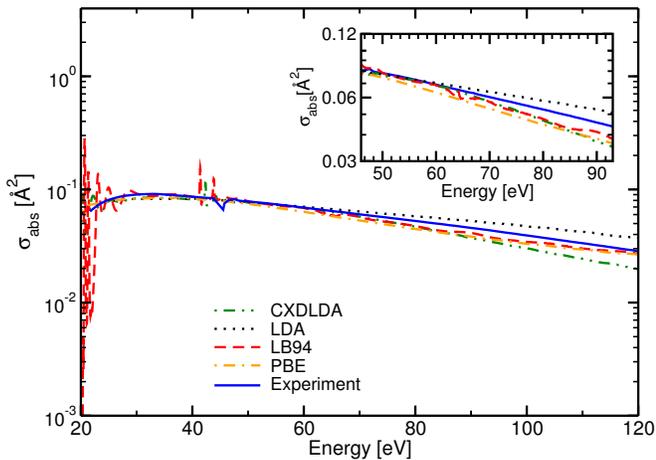}
\caption{Ne absorption cross-section (logarithmic scale) above the first ionization threshold.
Result for different TDDFT xc functionals, LDA (black), PBE (orange), CXD-LDA (green),  LB94 (red), 
compared with experimental data (blue)~\cite{Samson:2002hx}. 
In the inset we focus on the range of energies relevant for ionization from a $\omega=93$~eV laser
pulse.  }
\label{CROSS_NEON:fig}
\end{figure} 
We implement our boundary condition by inserting into the system's Hamiltonian 
an additional (spherically symmetric) imaginary potential $V_{\rm CAP}(r)$,  
acting at a certain distance $R_{\rm CAP}$ from the center of the box
of radius $R$,
\begin{equation}
	\label{eq:cap_pot} 
  V_{\rm CAP}(r)= -i\eta \left\{ 
	\begin{array}{ll}
		0 & \mbox{if $r < R_{\rm CAP}$} \\
		\sin^{2}\left( \frac{\pi(r-R_{\rm CAP})}{2(R-R_{\rm CAP})} \right) & \mbox{if $ R_{\rm CAP}\leq r \leq R$} 
	\end{array}
	\right. \,.
\end{equation}
Time propagation with a Hamiltonian containing $V_{\rm CAP}$, enforces 
a wavefunction damping in the region near the edges of the simulation box.
The absorption properties of this CAP as a function of the outgoing electron's kinetic energy, 
depends on the values of $\eta$ and $R_{\rm CAP}$. 

We find that for a spherical box of radius $R=16$~\AA~, a CAP having $\eta = 1$
and $R_{\rm CAP}=8$~\AA~, is enough to guarantee good continuum properties 
in the outgoing electron's kinetic energy range that we consider in this work for Ne and Ar. 
\new{To check this we have compared the experimental absorption spectra to the one obtained 
from TDDFT for different absorbing boxes, until reflections are negligible for this range.}

\begin{figure}
\includegraphics[width=\columnwidth]{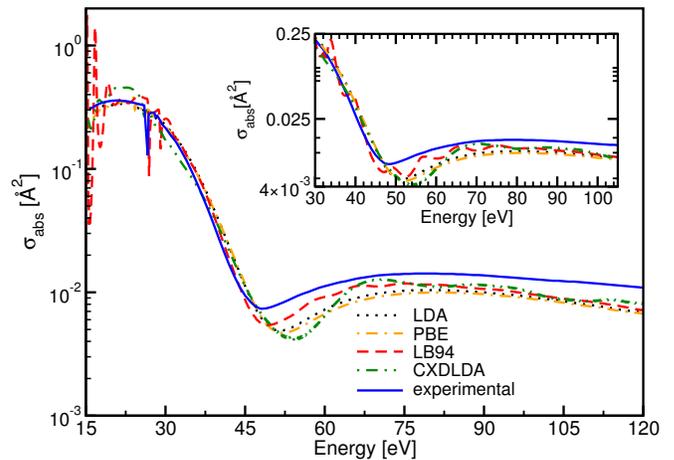}
\caption{
Ar absorption cross-section (logarithmic scale) above the first ionization threshold.
Result for different TDDFT xc functionals, LDA (black), PBE (orange),  CXD-LDA (green),  LB94 (red), 
compared with experimental data (blue)~\cite{Samson:2002hx}.
In the inset we focus on the range of energies relevant for ionization from a $\omega=105$~eV laser
pulse.
}
\label{CROSS_ARGON:fig}
\end{figure}

In order to assess the quality of the present choice, 
in Figs.~\ref{CROSS_NEON:fig} and \ref{CROSS_ARGON:fig}, 
we show Ne and Ar absorption cross-sections $\sigma$ obtained with different 
xc functionals 
for energies in the continuum, above the first ionization threshold. 
These cross-sections have been calculated in the linear regime analyzing the 
time evolution of the atomic dipole moment~\cite{Marques:2011ud}.

For both Ne and Ar, the cross-section presents spurious oscillations reminiscent of 
box states for energies $\lesssim 20$~eV above the ionization threshold. 
For larger values our CAP is well absorbing 
and the cross-sections smoothly follow the 
experimental ones~\cite{Samson:2002hx}. 

\begin{table*}
\begin{ruledtabular}
\caption{Ne and Ar pseudopotential vs all electron relative percentage errors for the 
outermost valence states for different xc functionals.}\label{TABLE_PSEUDO}
\begin{tabular}{cccccccc}
{\bf LDA Rel. err. (\%)} & 
$Ne^{+}$ / $Ar^{+}$ &
$Ne^{2+}$ / $Ar^{2+}$ &
$Ne^{3+}$ / $Ar^{3+}$ &
$Ne^{4+}$ / $Ar^{4+}$ &
$Ne^{5+}$ / $Ar^{5+}$ &
$Ne^{6+}$ / $Ar^{6+}$ &
$Ne^{7+}$ / $Ar^{7+}$ \\\hline
$2s$ / $3s$ &
-0.39 / 0.04 &
-0.41 / 0.10 &
-0.11 / 0.20 &
0.42 / 0.32 &
1.13 / 0.49 &
1.97 / 0.73 &
1.39 / 1.10 \\
$2p$ / $3p$ &
0.18 / 0.06 &
0.99 / 0.14 &
2.20 / 0.25 &
3.68 / 0.41 &
5.31 / 0.61 &
7.05 / 0.91 &
7.18 / 1.35 \\
$3d$ / $3d$ &
-0.07 / 0.08 & 
0.33 / 0.21 &
1.93 / 0.37 &
4.12 / 0.59 &
6.47 / 0.89 &
8.82 / 1.33 &
10.10 / 2.02 \\
{\bf PBE Rel. err. (\%)} \\
$2s$ / $3s$ &
-0.42 / 0.03 &
-0.47 / 0.06 &
-0.18 / 0.12 &
0.35 / 0.18 &
1.05 / 0.27 &
1.91 / 0.40 &
1.21 / 0.72 \\
$2p$ / $3p$ &
0.16 / 0.04 &
0.98 / 0.10 &
2.21 / 0.17 &
3.69 / 0.26 &
5.35 / 0.39 &
7.21 / 0.59 &
7.40 / 0.97 \\
$3d$ / $3d$ &
-0.08 / 0.08 & 
0.30 / 0.18 &
1.85 / 0.31 &
3.99 / 0.48 &
6.32 / 0.72 &
8.73 / 1.09 &
10.03 / 1.75 \\
{\bf LB94 Rel. err. (\%)} \\
$2s$ / $3s$ &
-0.19 / 0.10 &
-0.03 / 0.27 &
0.39 / 0.51 &
0.99 / 0.85 &
1.60 / 1.33 &
1.96 / 2.04 &
1.88 / 3.26 \\
$2p$ / $3p$ &
0.34 / 0.12 &
1.22 / 0.30 &
2.48 / 0.58 &
3.99 / 0.96 &
5.61 / 1.49 &
7.16 / 2.28 &
8.11 / 3.63 \\
$3d$ / $3d$ &
-0.10 / 0.08 & 
0.47 / 0.35 &
1.79 / 0.74 &
3.58 / 1.28 &
5.56 / 2.04 &
7.58 / 3.11 &
8.87 / 4.93 \\
{\bf CXD-LDA Rel. err. (\%)} \\
$2s$ / $3s$ &
-0.21 / 0.07 &
-0.41 / 0.10 &
0.02 / 0.06 &
0.59 / 0.16 &
1.28 / 0.53 &
1.89 / 0.60 &
1.71 / 1.65 \\
$2p$ / $3p$ &
0.35 / 0.09 &
0.84 / 0.14 &
2.15 / 0.09 &
3.61 / 0.22 &
5.19 / 0.65 &
6.67 / 0.75 &
7.33 / 1.91 \\
$3d$ / $3d$ &
0.04 / 0.06 & 
0.54 / 0.19 &
1.83 / 0.15 &
3.63 / 0.30 &
5.64 / 0.86 &
7.71 / 1.07 &
9.36 / 2.43 \\
\end{tabular}
\end{ruledtabular}
\end{table*} 

If we discard non-linear effects, the kinetic energy of an electron ejected 
by an ionizing laser pulse is given by $E = I_p-\omega$, 
where $I_p$ is the ionization potential of the bound electron and $\omega$ the laser energy.
Under the assumption that the energy absorbed from the laser is integrally transformed into 
kinetic energy of the escaping electron, we can conclude that the absorption
cross-section in the continuum is proportional to the electron photoemission probability.
The $I_p$ in TDDFT is given by the KS eigenvalue of each bound electron.
A rough estimate of the TDDFT quality attained in the description of ionization processes
initiated by a laser of a given frequency $\omega$, is therefore given by the behavior of
$\sigma$ in an energy range identified by the deeper and higher KS ionization potentials.
In the insets of Figs.~\ref{CROSS_NEON:fig} and \ref{CROSS_ARGON:fig} we plot $\sigma$ 
in the energy range relevant for a laser of $\omega=93$~eV (Ne) and $\omega=105$~eV (Ar)
respectively.

We can therefore conclude that in the energy range associated to our pulses,  
there are no spurious reflections and the absorption cross-sections 
agree remarkably with the experimental ones.

\subsection{Pseudopotential accuracy}\label{PSEUDOPOTENTIALS:sec}

In table~\ref{TABLE_PSEUDO}, we show the relative percentage errors introduced by the pseudopotential 
in the outermost valence energy levels of Ar and Ne, for increasing ionized species
and the different xc functionals tested. 
The error is here evaluated relative to an all-electron calculation.
Since our pseudopotentials have been generated from a \new{neutral} ground state configuration,
the errors increase linearly as a function of the charged state.
Here the errors are larger for Ne than for Ar, 
because Ne pseudopotentials have been generated using a larger radial cutoff. 
\new{The spacing of 0.16${\rm\AA}$ we have used is small enough to describe accurately the steep Coulomb potential
for the innermost core eigenvalues with charge $+6$ and $+7$. 
The eigenvalue errors for all the functionals are between 0.001 and 0.01 Ha.}


\begin{thebibliography}{44}%
\makeatletter
\providecommand \@ifxundefined [1]{%
 \@ifx{#1\undefined}
}%
\providecommand \@ifnum [1]{%
 \ifnum #1\expandafter \@firstoftwo
 \else \expandafter \@secondoftwo
 \fi
}%
\providecommand \@ifx [1]{%
 \ifx #1\expandafter \@firstoftwo
 \else \expandafter \@secondoftwo
 \fi
}%
\providecommand \natexlab [1]{#1}%
\providecommand \enquote  [1]{``#1''}%
\providecommand \bibnamefont  [1]{#1}%
\providecommand \bibfnamefont [1]{#1}%
\providecommand \citenamefont [1]{#1}%
\providecommand \href@noop [0]{\@secondoftwo}%
\providecommand \href [0]{\begingroup \@sanitize@url \@href}%
\providecommand \@href[1]{\@@startlink{#1}\@@href}%
\providecommand \@@href[1]{\endgroup#1\@@endlink}%
\providecommand \@sanitize@url [0]{\catcode `\\12\catcode `\$12\catcode
  `\&12\catcode `\#12\catcode `\^12\catcode `\_12\catcode `\%12\relax}%
\providecommand \@@startlink[1]{}%
\providecommand \@@endlink[0]{}%
\providecommand \url  [0]{\begingroup\@sanitize@url \@url }%
\providecommand \@url [1]{\endgroup\@href {#1}{\urlprefix }}%
\providecommand \urlprefix  [0]{URL }%
\providecommand \Eprint [0]{\href }%
\providecommand \doibase [0]{http://dx.doi.org/}%
\providecommand \selectlanguage [0]{\@gobble}%
\providecommand \bibinfo  [0]{\@secondoftwo}%
\providecommand \bibfield  [0]{\@secondoftwo}%
\providecommand \translation [1]{[#1]}%
\providecommand \BibitemOpen [0]{}%
\providecommand \bibitemStop [0]{}%
\providecommand \bibitemNoStop [0]{.\EOS\space}%
\providecommand \EOS [0]{\spacefactor3000\relax}%
\providecommand \BibitemShut  [1]{\csname bibitem#1\endcsname}%
\let\auto@bib@innerbib\@empty
\bibitem [{\citenamefont {Emma}\ \emph {et~al.}(2010)\citenamefont {Emma},
  \citenamefont {Akre}, \citenamefont {Arthur}, \citenamefont {Bionta},
  \citenamefont {Bostedt}, \citenamefont {Bozek}, \citenamefont {Brachmann},
  \citenamefont {Bucksbaum}, \citenamefont {Coffee}, \citenamefont {Decker},
  \citenamefont {Ding}, \citenamefont {Dowell}, \citenamefont {Edstrom},
  \citenamefont {Fisher}, \citenamefont {Frisch}, \citenamefont {Gilevich},
  \citenamefont {Hastings}, \citenamefont {Hays}, \citenamefont {Hering},
  \citenamefont {Huang}, \citenamefont {Iverson}, \citenamefont {Loos},
  \citenamefont {Messerschmidt}, \citenamefont {Miahnahri}, \citenamefont
  {Moeller}, \citenamefont {Nuhn}, \citenamefont {Pile}, \citenamefont
  {Ratner}, \citenamefont {Rzepiela}, \citenamefont {Schultz}, \citenamefont
  {Smith}, \citenamefont {Stefan}, \citenamefont {Tompkins}, \citenamefont
  {Turner}, \citenamefont {Welch}, \citenamefont {White}, \citenamefont {Wu},
  \citenamefont {Yocky},\ and\ \citenamefont {Galayda}}]{go1}%
  \BibitemOpen
  \bibfield  {author} {\bibinfo {author} {\bibfnamefont {P.}~\bibnamefont
  {Emma}}, \bibinfo {author} {\bibfnamefont {R.}~\bibnamefont {Akre}}, \bibinfo
  {author} {\bibfnamefont {J.}~\bibnamefont {Arthur}}, \bibinfo {author}
  {\bibfnamefont {R.}~\bibnamefont {Bionta}}, \bibinfo {author} {\bibfnamefont
  {C.}~\bibnamefont {Bostedt}}, \bibinfo {author} {\bibfnamefont
  {J.}~\bibnamefont {Bozek}}, \bibinfo {author} {\bibfnamefont
  {A.}~\bibnamefont {Brachmann}}, \bibinfo {author} {\bibfnamefont
  {P.}~\bibnamefont {Bucksbaum}}, \bibinfo {author} {\bibfnamefont
  {R.}~\bibnamefont {Coffee}}, \bibinfo {author} {\bibfnamefont {F.-J.}\
  \bibnamefont {Decker}}, \bibinfo {author} {\bibfnamefont {Y.}~\bibnamefont
  {Ding}}, \bibinfo {author} {\bibfnamefont {D.}~\bibnamefont {Dowell}},
  \bibinfo {author} {\bibfnamefont {S.}~\bibnamefont {Edstrom}}, \bibinfo
  {author} {\bibfnamefont {A.}~\bibnamefont {Fisher}}, \bibinfo {author}
  {\bibfnamefont {J.}~\bibnamefont {Frisch}}, \bibinfo {author} {\bibfnamefont
  {S.}~\bibnamefont {Gilevich}}, \bibinfo {author} {\bibfnamefont
  {J.}~\bibnamefont {Hastings}}, \bibinfo {author} {\bibfnamefont
  {G.}~\bibnamefont {Hays}}, \bibinfo {author} {\bibfnamefont {P.}~\bibnamefont
  {Hering}}, \bibinfo {author} {\bibfnamefont {Z.}~\bibnamefont {Huang}},
  \bibinfo {author} {\bibfnamefont {R.}~\bibnamefont {Iverson}}, \bibinfo
  {author} {\bibfnamefont {H.}~\bibnamefont {Loos}}, \bibinfo {author}
  {\bibfnamefont {M.}~\bibnamefont {Messerschmidt}}, \bibinfo {author}
  {\bibfnamefont {A.}~\bibnamefont {Miahnahri}}, \bibinfo {author}
  {\bibfnamefont {S.}~\bibnamefont {Moeller}}, \bibinfo {author} {\bibfnamefont
  {H.-D.}\ \bibnamefont {Nuhn}}, \bibinfo {author} {\bibfnamefont
  {G.}~\bibnamefont {Pile}}, \bibinfo {author} {\bibfnamefont {D.}~\bibnamefont
  {Ratner}}, \bibinfo {author} {\bibfnamefont {J.}~\bibnamefont {Rzepiela}},
  \bibinfo {author} {\bibfnamefont {D.}~\bibnamefont {Schultz}}, \bibinfo
  {author} {\bibfnamefont {T.}~\bibnamefont {Smith}}, \bibinfo {author}
  {\bibfnamefont {P.}~\bibnamefont {Stefan}}, \bibinfo {author} {\bibfnamefont
  {H.}~\bibnamefont {Tompkins}}, \bibinfo {author} {\bibfnamefont
  {J.}~\bibnamefont {Turner}}, \bibinfo {author} {\bibfnamefont
  {J.}~\bibnamefont {Welch}}, \bibinfo {author} {\bibfnamefont
  {W.}~\bibnamefont {White}}, \bibinfo {author} {\bibfnamefont
  {J.}~\bibnamefont {Wu}}, \bibinfo {author} {\bibfnamefont {G.}~\bibnamefont
  {Yocky}}, \ and\ \bibinfo {author} {\bibfnamefont {J.}~\bibnamefont
  {Galayda}},\ }\href@noop {} {\bibfield  {journal} {\bibinfo  {journal} {Nat.
  Photonics}\ }\textbf {\bibinfo {volume} {4}},\ \bibinfo {pages} {641}
  (\bibinfo {year} {2010})}\BibitemShut {NoStop}%
\bibitem [{\citenamefont {Ackermann}\ \emph {et~al.}(2007)\citenamefont
  {Ackermann}, \citenamefont {Asova}, \citenamefont {Ayvazyan}, \citenamefont
  {Azima}, \citenamefont {Baboi}, \citenamefont {Bähr}, \citenamefont
  {Balandin}, \citenamefont {Beutner}, \citenamefont {Brandt}, \citenamefont
  {Bolzmann}, \citenamefont {Brinkmann}, \citenamefont {Brovko}, \citenamefont
  {Castellano}, \citenamefont {Castro}, \citenamefont {Catani}, \citenamefont
  {Chiadroni}, \citenamefont {Choroba}, \citenamefont {Cianchi}, \citenamefont
  {Costello}, \citenamefont {Cubaynes}, \citenamefont {Dardis}, \citenamefont
  {Decking}, \citenamefont {Delsim-Hashemi}, \citenamefont {Delserieys},
  \citenamefont {Pirro}, \citenamefont {Dohlus}, \citenamefont {Düsterer},
  \citenamefont {Eckhardt}, \citenamefont {Edwards}, \citenamefont {Faatz},
  \citenamefont {Feldhaus}, \citenamefont {Flöttmann}, \citenamefont {Frisch},
  \citenamefont {Fröhlich}, \citenamefont {Garvey}, \citenamefont {Gensch},
  \citenamefont {Gerth}, \citenamefont {Görler}, \citenamefont {Golubeva},
  \citenamefont {Grabosch}, \citenamefont {Grecki}, \citenamefont {Grimm},
  \citenamefont {Hacker}, \citenamefont {Hahn}, \citenamefont {Han},
  \citenamefont {Honkavaara}, \citenamefont {Hott}, \citenamefont {Hüning},
  \citenamefont {Ivanisenko}, \citenamefont {Jaeschke}, \citenamefont
  {Jalmuzna}, \citenamefont {Jezynski}, \citenamefont {Kammering},
  \citenamefont {Katalev}, \citenamefont {Kavanagh}, \citenamefont {Kennedy},
  \citenamefont {Khodyachykh}, \citenamefont {Klose}, \citenamefont
  {Kocharyan}, \citenamefont {Körfer}, \citenamefont {Kollewe}, \citenamefont
  {Koprek}, \citenamefont {Korepanov}, \citenamefont {Kostin}, \citenamefont
  {Krassilnikov}, \citenamefont {Kube}, \citenamefont {Kuhlmann}, \citenamefont
  {Lewis}, \citenamefont {Lilje}, \citenamefont {Limberg}, \citenamefont
  {Lipka}, \citenamefont {Löhl}, \citenamefont {Luna}, \citenamefont {Luong},
  \citenamefont {Martins}, \citenamefont {Meyer}, \citenamefont {Michelato},
  \citenamefont {Miltchev}, \citenamefont {Möller}, \citenamefont {Monaco},
  \citenamefont {Müller}, \citenamefont {Napieralski}, \citenamefont {Napoly},
  \citenamefont {Nicolosi}, \citenamefont {Nölle}, \citenamefont {Nuñez},
  \citenamefont {Oppelt}, \citenamefont {Pagani}, \citenamefont {Paparella},
  \citenamefont {Pchalek}, \citenamefont {Pedregosa-Gutierrez}, \citenamefont
  {Petersen}, \citenamefont {Petrosyan}, \citenamefont {Petrosyan},
  \citenamefont {Pflüger}, \citenamefont {Plönjes}, \citenamefont {Poletto},
  \citenamefont {Pozniak}, \citenamefont {Prat}, \citenamefont {Proch},
  \citenamefont {Pucyk}, \citenamefont {Radcliffe}, \citenamefont {Redlin},
  \citenamefont {Rehlich}, \citenamefont {Richter}, \citenamefont {Roehrs},
  \citenamefont {Roensch}, \citenamefont {Romaniuk}, \citenamefont {Ross},
  \citenamefont {Rossbach}, \citenamefont {Rybnikov}, \citenamefont {Sachwitz},
  \citenamefont {Saldin}, \citenamefont {Sandner}, \citenamefont {Schlarb},
  \citenamefont {Schmidt}, \citenamefont {Schmitz}, \citenamefont {Schmüser},
  \citenamefont {Schneider}, \citenamefont {Schneidmiller}, \citenamefont
  {Schnepp}, \citenamefont {Schreiber}, \citenamefont {Seidel}, \citenamefont
  {Sertore}, \citenamefont {Shabunov}, \citenamefont {Simon}, \citenamefont
  {Simrock}, \citenamefont {Sombrowski}, \citenamefont {Sorokin}, \citenamefont
  {Spanknebel}, \citenamefont {Spesyvtsev}, \citenamefont {Staykov},
  \citenamefont {Steffen}, \citenamefont {Stephan}, \citenamefont {Stulle},
  \citenamefont {Thom}, \citenamefont {Tiedtke}, \citenamefont {Tischer},
  \citenamefont {Toleikis}, \citenamefont {Treusch}, \citenamefont {Trines},
  \citenamefont {Tsakov}, \citenamefont {Vogel}, \citenamefont {Weiland},
  \citenamefont {Weise}, \citenamefont {Wellhöfer}, \citenamefont {Wendt},
  \citenamefont {Will}, \citenamefont {Winter}, \citenamefont {Wittenburg},
  \citenamefont {Wurth}, \citenamefont {Yeates}, \citenamefont {Yurkov},
  \citenamefont {Zagorodnov},\ and\ \citenamefont {Zapfe}}]{1}%
  \BibitemOpen
  \bibfield  {author} {\bibinfo {author} {\bibfnamefont {W.}~\bibnamefont
  {Ackermann}}, \bibinfo {author} {\bibfnamefont {G.}~\bibnamefont {Asova}},
  \bibinfo {author} {\bibfnamefont {V.}~\bibnamefont {Ayvazyan}}, \bibinfo
  {author} {\bibfnamefont {A.}~\bibnamefont {Azima}}, \bibinfo {author}
  {\bibfnamefont {N.}~\bibnamefont {Baboi}}, \bibinfo {author} {\bibfnamefont
  {J.}~\bibnamefont {Bähr}}, \bibinfo {author} {\bibfnamefont
  {V.}~\bibnamefont {Balandin}}, \bibinfo {author} {\bibfnamefont
  {B.}~\bibnamefont {Beutner}}, \bibinfo {author} {\bibfnamefont
  {A.}~\bibnamefont {Brandt}}, \bibinfo {author} {\bibfnamefont
  {A.}~\bibnamefont {Bolzmann}}, \bibinfo {author} {\bibfnamefont
  {R.}~\bibnamefont {Brinkmann}}, \bibinfo {author} {\bibfnamefont {O.~I.}\
  \bibnamefont {Brovko}}, \bibinfo {author} {\bibfnamefont {M.}~\bibnamefont
  {Castellano}}, \bibinfo {author} {\bibfnamefont {P.}~\bibnamefont {Castro}},
  \bibinfo {author} {\bibfnamefont {L.}~\bibnamefont {Catani}}, \bibinfo
  {author} {\bibfnamefont {E.}~\bibnamefont {Chiadroni}}, \bibinfo {author}
  {\bibfnamefont {S.}~\bibnamefont {Choroba}}, \bibinfo {author} {\bibfnamefont
  {A.}~\bibnamefont {Cianchi}}, \bibinfo {author} {\bibfnamefont {J.~T.}\
  \bibnamefont {Costello}}, \bibinfo {author} {\bibfnamefont {D.}~\bibnamefont
  {Cubaynes}}, \bibinfo {author} {\bibfnamefont {J.}~\bibnamefont {Dardis}},
  \bibinfo {author} {\bibfnamefont {W.}~\bibnamefont {Decking}}, \bibinfo
  {author} {\bibfnamefont {H.}~\bibnamefont {Delsim-Hashemi}}, \bibinfo
  {author} {\bibfnamefont {A.}~\bibnamefont {Delserieys}}, \bibinfo {author}
  {\bibfnamefont {G.~D.}\ \bibnamefont {Pirro}}, \bibinfo {author}
  {\bibfnamefont {M.}~\bibnamefont {Dohlus}}, \bibinfo {author} {\bibfnamefont
  {S.}~\bibnamefont {Düsterer}}, \bibinfo {author} {\bibfnamefont
  {A.}~\bibnamefont {Eckhardt}}, \bibinfo {author} {\bibfnamefont {H.~T.}\
  \bibnamefont {Edwards}}, \bibinfo {author} {\bibfnamefont {B.}~\bibnamefont
  {Faatz}}, \bibinfo {author} {\bibfnamefont {J.}~\bibnamefont {Feldhaus}},
  \bibinfo {author} {\bibfnamefont {K.}~\bibnamefont {Flöttmann}}, \bibinfo
  {author} {\bibfnamefont {J.}~\bibnamefont {Frisch}}, \bibinfo {author}
  {\bibfnamefont {L.}~\bibnamefont {Fröhlich}}, \bibinfo {author}
  {\bibfnamefont {T.}~\bibnamefont {Garvey}}, \bibinfo {author} {\bibfnamefont
  {U.}~\bibnamefont {Gensch}}, \bibinfo {author} {\bibfnamefont
  {C.}~\bibnamefont {Gerth}}, \bibinfo {author} {\bibfnamefont
  {M.}~\bibnamefont {Görler}}, \bibinfo {author} {\bibfnamefont
  {N.}~\bibnamefont {Golubeva}}, \bibinfo {author} {\bibfnamefont {H.-J.}\
  \bibnamefont {Grabosch}}, \bibinfo {author} {\bibfnamefont {M.}~\bibnamefont
  {Grecki}}, \bibinfo {author} {\bibfnamefont {O.}~\bibnamefont {Grimm}},
  \bibinfo {author} {\bibfnamefont {K.}~\bibnamefont {Hacker}}, \bibinfo
  {author} {\bibfnamefont {U.}~\bibnamefont {Hahn}}, \bibinfo {author}
  {\bibfnamefont {J.~H.}\ \bibnamefont {Han}}, \bibinfo {author} {\bibfnamefont
  {K.}~\bibnamefont {Honkavaara}}, \bibinfo {author} {\bibfnamefont
  {T.}~\bibnamefont {Hott}}, \bibinfo {author} {\bibfnamefont {M.}~\bibnamefont
  {Hüning}}, \bibinfo {author} {\bibfnamefont {Y.}~\bibnamefont {Ivanisenko}},
  \bibinfo {author} {\bibfnamefont {E.}~\bibnamefont {Jaeschke}}, \bibinfo
  {author} {\bibfnamefont {W.}~\bibnamefont {Jalmuzna}}, \bibinfo {author}
  {\bibfnamefont {T.}~\bibnamefont {Jezynski}}, \bibinfo {author}
  {\bibfnamefont {R.}~\bibnamefont {Kammering}}, \bibinfo {author}
  {\bibfnamefont {V.}~\bibnamefont {Katalev}}, \bibinfo {author} {\bibfnamefont
  {K.}~\bibnamefont {Kavanagh}}, \bibinfo {author} {\bibfnamefont {E.~T.}\
  \bibnamefont {Kennedy}}, \bibinfo {author} {\bibfnamefont {S.}~\bibnamefont
  {Khodyachykh}}, \bibinfo {author} {\bibfnamefont {K.}~\bibnamefont {Klose}},
  \bibinfo {author} {\bibfnamefont {V.}~\bibnamefont {Kocharyan}}, \bibinfo
  {author} {\bibfnamefont {M.}~\bibnamefont {Körfer}}, \bibinfo {author}
  {\bibfnamefont {M.}~\bibnamefont {Kollewe}}, \bibinfo {author} {\bibfnamefont
  {W.}~\bibnamefont {Koprek}}, \bibinfo {author} {\bibfnamefont
  {S.}~\bibnamefont {Korepanov}}, \bibinfo {author} {\bibfnamefont
  {D.}~\bibnamefont {Kostin}}, \bibinfo {author} {\bibfnamefont
  {M.}~\bibnamefont {Krassilnikov}}, \bibinfo {author} {\bibfnamefont
  {G.}~\bibnamefont {Kube}}, \bibinfo {author} {\bibfnamefont {M.}~\bibnamefont
  {Kuhlmann}}, \bibinfo {author} {\bibfnamefont {C.~L.~S.}\ \bibnamefont
  {Lewis}}, \bibinfo {author} {\bibfnamefont {L.}~\bibnamefont {Lilje}},
  \bibinfo {author} {\bibfnamefont {T.}~\bibnamefont {Limberg}}, \bibinfo
  {author} {\bibfnamefont {D.}~\bibnamefont {Lipka}}, \bibinfo {author}
  {\bibfnamefont {F.}~\bibnamefont {Löhl}}, \bibinfo {author} {\bibfnamefont
  {H.}~\bibnamefont {Luna}}, \bibinfo {author} {\bibfnamefont {M.}~\bibnamefont
  {Luong}}, \bibinfo {author} {\bibfnamefont {M.}~\bibnamefont {Martins}},
  \bibinfo {author} {\bibfnamefont {M.}~\bibnamefont {Meyer}}, \bibinfo
  {author} {\bibfnamefont {P.}~\bibnamefont {Michelato}}, \bibinfo {author}
  {\bibfnamefont {V.}~\bibnamefont {Miltchev}}, \bibinfo {author}
  {\bibfnamefont {W.~D.}\ \bibnamefont {Möller}}, \bibinfo {author}
  {\bibfnamefont {L.}~\bibnamefont {Monaco}}, \bibinfo {author} {\bibfnamefont
  {W.~F.~O.}\ \bibnamefont {Müller}}, \bibinfo {author} {\bibfnamefont
  {O.}~\bibnamefont {Napieralski}}, \bibinfo {author} {\bibfnamefont
  {O.}~\bibnamefont {Napoly}}, \bibinfo {author} {\bibfnamefont
  {P.}~\bibnamefont {Nicolosi}}, \bibinfo {author} {\bibfnamefont
  {D.}~\bibnamefont {Nölle}}, \bibinfo {author} {\bibfnamefont
  {T.}~\bibnamefont {Nuñez}}, \bibinfo {author} {\bibfnamefont
  {A.}~\bibnamefont {Oppelt}}, \bibinfo {author} {\bibfnamefont
  {C.}~\bibnamefont {Pagani}}, \bibinfo {author} {\bibfnamefont
  {R.}~\bibnamefont {Paparella}}, \bibinfo {author} {\bibfnamefont
  {N.}~\bibnamefont {Pchalek}}, \bibinfo {author} {\bibfnamefont
  {J.}~\bibnamefont {Pedregosa-Gutierrez}}, \bibinfo {author} {\bibfnamefont
  {B.}~\bibnamefont {Petersen}}, \bibinfo {author} {\bibfnamefont
  {G.}~\bibnamefont {Petrosyan}}, \bibinfo {author} {\bibfnamefont
  {L.}~\bibnamefont {Petrosyan}}, \bibinfo {author} {\bibfnamefont
  {J.}~\bibnamefont {Pflüger}}, \bibinfo {author} {\bibfnamefont
  {E.}~\bibnamefont {Plönjes}}, \bibinfo {author} {\bibfnamefont
  {L.}~\bibnamefont {Poletto}}, \bibinfo {author} {\bibfnamefont
  {K.}~\bibnamefont {Pozniak}}, \bibinfo {author} {\bibfnamefont
  {E.}~\bibnamefont {Prat}}, \bibinfo {author} {\bibfnamefont {D.}~\bibnamefont
  {Proch}}, \bibinfo {author} {\bibfnamefont {P.}~\bibnamefont {Pucyk}},
  \bibinfo {author} {\bibfnamefont {P.}~\bibnamefont {Radcliffe}}, \bibinfo
  {author} {\bibfnamefont {H.}~\bibnamefont {Redlin}}, \bibinfo {author}
  {\bibfnamefont {K.}~\bibnamefont {Rehlich}}, \bibinfo {author} {\bibfnamefont
  {M.}~\bibnamefont {Richter}}, \bibinfo {author} {\bibfnamefont
  {M.}~\bibnamefont {Roehrs}}, \bibinfo {author} {\bibfnamefont
  {J.}~\bibnamefont {Roensch}}, \bibinfo {author} {\bibfnamefont
  {R.}~\bibnamefont {Romaniuk}}, \bibinfo {author} {\bibfnamefont
  {M.}~\bibnamefont {Ross}}, \bibinfo {author} {\bibfnamefont {J.}~\bibnamefont
  {Rossbach}}, \bibinfo {author} {\bibfnamefont {V.}~\bibnamefont {Rybnikov}},
  \bibinfo {author} {\bibfnamefont {M.}~\bibnamefont {Sachwitz}}, \bibinfo
  {author} {\bibfnamefont {E.~L.}\ \bibnamefont {Saldin}}, \bibinfo {author}
  {\bibfnamefont {W.}~\bibnamefont {Sandner}}, \bibinfo {author} {\bibfnamefont
  {H.}~\bibnamefont {Schlarb}}, \bibinfo {author} {\bibfnamefont
  {B.}~\bibnamefont {Schmidt}}, \bibinfo {author} {\bibfnamefont
  {M.}~\bibnamefont {Schmitz}}, \bibinfo {author} {\bibfnamefont
  {P.}~\bibnamefont {Schmüser}}, \bibinfo {author} {\bibfnamefont {J.~R.}\
  \bibnamefont {Schneider}}, \bibinfo {author} {\bibfnamefont {E.~A.}\
  \bibnamefont {Schneidmiller}}, \bibinfo {author} {\bibfnamefont
  {S.}~\bibnamefont {Schnepp}}, \bibinfo {author} {\bibfnamefont
  {S.}~\bibnamefont {Schreiber}}, \bibinfo {author} {\bibfnamefont
  {M.}~\bibnamefont {Seidel}}, \bibinfo {author} {\bibfnamefont
  {D.}~\bibnamefont {Sertore}}, \bibinfo {author} {\bibfnamefont {A.~V.}\
  \bibnamefont {Shabunov}}, \bibinfo {author} {\bibfnamefont {C.}~\bibnamefont
  {Simon}}, \bibinfo {author} {\bibfnamefont {S.}~\bibnamefont {Simrock}},
  \bibinfo {author} {\bibfnamefont {E.}~\bibnamefont {Sombrowski}}, \bibinfo
  {author} {\bibfnamefont {A.}~\bibnamefont {Sorokin}}, \bibinfo {author}
  {\bibfnamefont {P.}~\bibnamefont {Spanknebel}}, \bibinfo {author}
  {\bibfnamefont {R.}~\bibnamefont {Spesyvtsev}}, \bibinfo {author}
  {\bibfnamefont {L.}~\bibnamefont {Staykov}}, \bibinfo {author} {\bibfnamefont
  {B.}~\bibnamefont {Steffen}}, \bibinfo {author} {\bibfnamefont
  {F.}~\bibnamefont {Stephan}}, \bibinfo {author} {\bibfnamefont
  {H.}~\bibnamefont {Stulle}}, \bibinfo {author} {\bibfnamefont
  {K.}~\bibnamefont {Thom}}, \bibinfo {author} {\bibfnamefont {M.}~\bibnamefont
  {Tiedtke}}, \bibinfo {author} {\bibfnamefont {S.}~\bibnamefont {Tischer}},
  \bibinfo {author} {\bibfnamefont {R.}~\bibnamefont {Toleikis}}, \bibinfo
  {author} {\bibfnamefont {D.}~\bibnamefont {Treusch}}, \bibinfo {author}
  {\bibfnamefont {I.}~\bibnamefont {Trines}}, \bibinfo {author} {\bibfnamefont
  {E.}~\bibnamefont {Tsakov}}, \bibinfo {author} {\bibfnamefont
  {T.}~\bibnamefont {Vogel}}, \bibinfo {author} {\bibfnamefont
  {T.}~\bibnamefont {Weiland}}, \bibinfo {author} {\bibfnamefont
  {H.}~\bibnamefont {Weise}}, \bibinfo {author} {\bibfnamefont
  {M.}~\bibnamefont {Wellhöfer}}, \bibinfo {author} {\bibfnamefont
  {M.}~\bibnamefont {Wendt}}, \bibinfo {author} {\bibfnamefont
  {I.}~\bibnamefont {Will}}, \bibinfo {author} {\bibfnamefont {A.}~\bibnamefont
  {Winter}}, \bibinfo {author} {\bibfnamefont {K.}~\bibnamefont {Wittenburg}},
  \bibinfo {author} {\bibfnamefont {W.}~\bibnamefont {Wurth}}, \bibinfo
  {author} {\bibfnamefont {P.}~\bibnamefont {Yeates}}, \bibinfo {author}
  {\bibfnamefont {M.~V.}\ \bibnamefont {Yurkov}}, \bibinfo {author}
  {\bibfnamefont {I.}~\bibnamefont {Zagorodnov}}, \ and\ \bibinfo {author}
  {\bibfnamefont {Z.}~\bibnamefont {Zapfe}},\ }\href@noop {} {\bibfield
  {journal} {\bibinfo  {journal} {Nat. Photonics}\ }\textbf {\bibinfo {volume}
  {1}},\ \bibinfo {pages} {336} (\bibinfo {year} {2007})}\BibitemShut {NoStop}%
\bibitem [{\citenamefont {Feldhaus}\ \emph {et~al.}(2013)\citenamefont
  {Feldhaus}, \citenamefont {Krikunova}, \citenamefont {Meyer}, \citenamefont
  {Möller}, \citenamefont {Moshammer}, \citenamefont {Rudenko}, \citenamefont
  {Tschentscher},\ and\ \citenamefont {Ullrich}}]{AND_MORE}%
  \BibitemOpen
  \bibfield  {author} {\bibinfo {author} {\bibfnamefont {J.}~\bibnamefont
  {Feldhaus}}, \bibinfo {author} {\bibfnamefont {M.}~\bibnamefont {Krikunova}},
  \bibinfo {author} {\bibfnamefont {M.}~\bibnamefont {Meyer}}, \bibinfo
  {author} {\bibfnamefont {T.}~\bibnamefont {Möller}}, \bibinfo {author}
  {\bibfnamefont {R.}~\bibnamefont {Moshammer}}, \bibinfo {author}
  {\bibfnamefont {A.}~\bibnamefont {Rudenko}}, \bibinfo {author} {\bibfnamefont
  {T.}~\bibnamefont {Tschentscher}}, \ and\ \bibinfo {author} {\bibfnamefont
  {J.}~\bibnamefont {Ullrich}},\ }\href
  {http://stacks.iop.org/0953-4075/46/i=16/a=164002} {\bibfield  {journal}
  {\bibinfo  {journal} {J. Phys. B: At. Mol. Opt. Phys.}\ }\textbf {\bibinfo
  {volume} {46}},\ \bibinfo {pages} {164002} (\bibinfo {year}
  {2013})}\BibitemShut {NoStop}%
\bibitem [{\citenamefont {Bostedt}\ \emph {et~al.}(2013)\citenamefont
  {Bostedt}, \citenamefont {Bozek}, \citenamefont {Bucksbaum}, \citenamefont
  {Coffee}, \citenamefont {Hastings}, \citenamefont {Huang}, \citenamefont
  {Lee}, \citenamefont {Schorb}, \citenamefont {Corlett}, \citenamefont
  {Denes}, \citenamefont {Emma}, \citenamefont {Falcone}, \citenamefont
  {Schoenlein}, \citenamefont {Doumy}, \citenamefont {Kanter}, \citenamefont
  {Kraessig}, \citenamefont {Southworth}, \citenamefont {Young}, \citenamefont
  {Fang}, \citenamefont {Hoener}, \citenamefont {Berrah}, \citenamefont
  {Roedig},\ and\ \citenamefont {DiMauro}}]{LAM_AGAIN_AGAIN}%
  \BibitemOpen
  \bibfield  {author} {\bibinfo {author} {\bibfnamefont {C.}~\bibnamefont
  {Bostedt}}, \bibinfo {author} {\bibfnamefont {J.~D.}\ \bibnamefont {Bozek}},
  \bibinfo {author} {\bibfnamefont {P.~H.}\ \bibnamefont {Bucksbaum}}, \bibinfo
  {author} {\bibfnamefont {R.~N.}\ \bibnamefont {Coffee}}, \bibinfo {author}
  {\bibfnamefont {J.~B.}\ \bibnamefont {Hastings}}, \bibinfo {author}
  {\bibfnamefont {Z.}~\bibnamefont {Huang}}, \bibinfo {author} {\bibfnamefont
  {R.~W.}\ \bibnamefont {Lee}}, \bibinfo {author} {\bibfnamefont
  {S.}~\bibnamefont {Schorb}}, \bibinfo {author} {\bibfnamefont {J.~N.}\
  \bibnamefont {Corlett}}, \bibinfo {author} {\bibfnamefont {P.}~\bibnamefont
  {Denes}}, \bibinfo {author} {\bibfnamefont {P.}~\bibnamefont {Emma}},
  \bibinfo {author} {\bibfnamefont {R.~W.}\ \bibnamefont {Falcone}}, \bibinfo
  {author} {\bibfnamefont {R.~W.}\ \bibnamefont {Schoenlein}}, \bibinfo
  {author} {\bibfnamefont {G.}~\bibnamefont {Doumy}}, \bibinfo {author}
  {\bibfnamefont {E.~P.}\ \bibnamefont {Kanter}}, \bibinfo {author}
  {\bibfnamefont {B.}~\bibnamefont {Kraessig}}, \bibinfo {author}
  {\bibfnamefont {S.}~\bibnamefont {Southworth}}, \bibinfo {author}
  {\bibfnamefont {L.}~\bibnamefont {Young}}, \bibinfo {author} {\bibfnamefont
  {L.}~\bibnamefont {Fang}}, \bibinfo {author} {\bibfnamefont {M.}~\bibnamefont
  {Hoener}}, \bibinfo {author} {\bibfnamefont {N.}~\bibnamefont {Berrah}},
  \bibinfo {author} {\bibfnamefont {C.}~\bibnamefont {Roedig}}, \ and\ \bibinfo
  {author} {\bibfnamefont {L.~F.}\ \bibnamefont {DiMauro}},\ }\href {\doibase
  doi:10.1088/0953-4075/46/16/164003} {\bibfield  {journal} {\bibinfo
  {journal} {J. Phys. B: At. Mol. Opt. Phys.}\ }\textbf {\bibinfo {volume}
  {46}},\ \bibinfo {pages} {164003} (\bibinfo {year} {2013})}\BibitemShut
  {NoStop}%
\bibitem [{\citenamefont {Sorokin}\ \emph {et~al.}(2007)\citenamefont
  {Sorokin}, \citenamefont {Bobashev}, \citenamefont {Feigl}, \citenamefont
  {Tiedtke}, \citenamefont {Wabnitz},\ and\ \citenamefont {Richter}}]{2}%
  \BibitemOpen
  \bibfield  {author} {\bibinfo {author} {\bibfnamefont {A.~A.}\ \bibnamefont
  {Sorokin}}, \bibinfo {author} {\bibfnamefont {S.~V.}\ \bibnamefont
  {Bobashev}}, \bibinfo {author} {\bibfnamefont {T.}~\bibnamefont {Feigl}},
  \bibinfo {author} {\bibfnamefont {K.}~\bibnamefont {Tiedtke}}, \bibinfo
  {author} {\bibfnamefont {H.}~\bibnamefont {Wabnitz}}, \ and\ \bibinfo
  {author} {\bibfnamefont {M.}~\bibnamefont {Richter}},\ }\href {\doibase
  10.1103/PhysRevLett.99.213002} {\bibfield  {journal} {\bibinfo  {journal}
  {Phys. Rev. Lett.}\ }\textbf {\bibinfo {volume} {99}},\ \bibinfo {pages}
  {213002} (\bibinfo {year} {2007})}\BibitemShut {NoStop}%
\bibitem [{\citenamefont {{M. Richter, S. V. Bobashev, A. A. Sorokin, K.
  Tiedtke}}(2010)}]{21}%
  \BibitemOpen
  \bibfield  {author} {\bibinfo {author} {\bibnamefont {{M. Richter, S. V.
  Bobashev, A. A. Sorokin, K. Tiedtke}}},\ }\href@noop {} {\bibfield  {journal}
  {\bibinfo  {journal} {J. Phys. B: At. Mol. Opt. Phys.}\ }\textbf {\bibinfo
  {volume} {43}},\ \bibinfo {pages} {194005} (\bibinfo {year}
  {2010})}\BibitemShut {NoStop}%
\bibitem [{\citenamefont {{M. G. Makris and P. Lambropoulos and A.
  Mihelic}}(2009)}]{3}%
  \BibitemOpen
  \bibfield  {author} {\bibinfo {author} {\bibnamefont {{M. G. Makris and P.
  Lambropoulos and A. Mihelic}}},\ }\href@noop {} {\bibfield  {journal}
  {\bibinfo  {journal} {Phys. Rev. Lett.}\ }\textbf {\bibinfo {volume} {102}},\
  \bibinfo {pages} {033002} (\bibinfo {year} {2009})}\BibitemShut {NoStop}%
\bibitem [{\citenamefont {{P. Lambropoulos and K. G. Papamihail and P.
  Decleva}}(2011)}]{4}%
  \BibitemOpen
  \bibfield  {author} {\bibinfo {author} {\bibnamefont {{P. Lambropoulos and K.
  G. Papamihail and P. Decleva}}},\ }\href@noop {} {\bibfield  {journal}
  {\bibinfo  {journal} {J. Phys. B: At. Mol. Opt. Phys.}\ }\textbf {\bibinfo
  {volume} {44}},\ \bibinfo {pages} {175402} (\bibinfo {year}
  {2011})}\BibitemShut {NoStop}%
\bibitem [{\citenamefont {Lambropoulos}\ and\ \citenamefont
  {Nikolopoulos}(2013)}]{41}%
  \BibitemOpen
  \bibfield  {author} {\bibinfo {author} {\bibfnamefont {P.}~\bibnamefont
  {Lambropoulos}}\ and\ \bibinfo {author} {\bibfnamefont {G.~M.}\ \bibnamefont
  {Nikolopoulos}},\ }\href@noop {} {\bibfield  {journal} {\bibinfo  {journal}
  {Eur. Phys. J. Special Topics}\ }\textbf {\bibinfo {volume} {222}},\ \bibinfo
  {pages} {2067} (\bibinfo {year} {2013})}\BibitemShut {NoStop}%
\bibitem [{\citenamefont {{P. Lambropoulos, G. M. Nikolopoulos, and K. G.
  Papamihail}}(2011)}]{LAMBRO}%
  \BibitemOpen
  \bibfield  {author} {\bibinfo {author} {\bibnamefont {{P. Lambropoulos, G. M.
  Nikolopoulos, and K. G. Papamihail}}},\ }\href@noop {} {\bibfield  {journal}
  {\bibinfo  {journal} {Phys. Rev. A}\ }\textbf {\bibinfo {volume} {83}},\
  \bibinfo {pages} {021407(R)} (\bibinfo {year} {2011})}\BibitemShut {NoStop}%
\bibitem [{\citenamefont {{G. M. Nikolopoulos and P.
  Lambropoulos}}(2014)}]{LAMBRO1}%
  \BibitemOpen
  \bibfield  {author} {\bibinfo {author} {\bibnamefont {{G. M. Nikolopoulos and
  P. Lambropoulos}}},\ }\href@noop {} {\bibfield  {journal} {\bibinfo
  {journal} {J. Phys. B: At. Mol. Opt. Phys.}\ }\textbf {\bibinfo {volume}
  {47}},\ \bibinfo {pages} {115001} (\bibinfo {year} {2014})}\BibitemShut
  {NoStop}%
\bibitem [{\citenamefont {Lambropoulos}\ \emph {et~al.}(1998)\citenamefont
  {Lambropoulos}, \citenamefont {Maragakis},\ and\ \citenamefont
  {Zhang}}]{LAM_AGAIN}%
  \BibitemOpen
  \bibfield  {author} {\bibinfo {author} {\bibfnamefont {P.}~\bibnamefont
  {Lambropoulos}}, \bibinfo {author} {\bibfnamefont {P.}~\bibnamefont
  {Maragakis}}, \ and\ \bibinfo {author} {\bibfnamefont {J.}~\bibnamefont
  {Zhang}},\ }\href {\doibase http://dx.doi.org/10.1016/S0370-1573(98)00027-1}
  {\bibfield  {journal} {\bibinfo  {journal} {Phys. Rep.}\ }\textbf {\bibinfo
  {volume} {305}},\ \bibinfo {pages} {203 } (\bibinfo {year}
  {1998})}\BibitemShut {NoStop}%
\bibitem [{\citenamefont {Rudek}\ \emph {et~al.}(2012)\citenamefont {Rudek},
  \citenamefont {Son}, \citenamefont {Foucar}, \citenamefont {Epp},
  \citenamefont {Erk}, \citenamefont {Hartmann}, \citenamefont {Adolph},
  \citenamefont {Andritschke}, \citenamefont {Aquila}, \citenamefont {Berrah},
  \citenamefont {Bostedt}, \citenamefont {Bozek}, \citenamefont {Coppola},
  \citenamefont {Filsinger}, \citenamefont {Gorke}, \citenamefont {Gorkhover},
  \citenamefont {Graafsma}, \citenamefont {Gumprecht}, \citenamefont
  {Hartmann}, \citenamefont {Hauser}, \citenamefont {Herrmann}, \citenamefont
  {Hirsemann}, \citenamefont {Holl}, \citenamefont {Hömke}, \citenamefont
  {Journel}, \citenamefont {Kaiser}, \citenamefont {Kimmel}, \citenamefont
  {Krasniqi}, \citenamefont {Kühnel}, \citenamefont {Matysek}, \citenamefont
  {Messerschmidt}, \citenamefont {Miesner}, \citenamefont {Möller},
  \citenamefont {Moshammer}, \citenamefont {Nagaya}, \citenamefont {Nilsson},
  \citenamefont {Potdevin}, \citenamefont {Pietschner}, \citenamefont {Reich},
  \citenamefont {Rupp}, \citenamefont {Schaller}, \citenamefont {Schlichting},
  \citenamefont {Schmidt}, \citenamefont {Schopper}, \citenamefont {Schorb},
  \citenamefont {Schröter}, \citenamefont {Schulz}, \citenamefont {Simon},
  \citenamefont {Soltau}, \citenamefont {Strüder}, \citenamefont {Ueda},
  \citenamefont {Weidenspointner}, \citenamefont {Santra}, \citenamefont
  {Ullrich}, \citenamefont {Rudenko},\ and\ \citenamefont
  {Rolles}}]{NATURE_PHOT}%
  \BibitemOpen
  \bibfield  {author} {\bibinfo {author} {\bibfnamefont {B.}~\bibnamefont
  {Rudek}}, \bibinfo {author} {\bibfnamefont {S.-K.}\ \bibnamefont {Son}},
  \bibinfo {author} {\bibfnamefont {L.}~\bibnamefont {Foucar}}, \bibinfo
  {author} {\bibfnamefont {S.~W.}\ \bibnamefont {Epp}}, \bibinfo {author}
  {\bibfnamefont {B.}~\bibnamefont {Erk}}, \bibinfo {author} {\bibfnamefont
  {R.}~\bibnamefont {Hartmann}}, \bibinfo {author} {\bibfnamefont
  {M.}~\bibnamefont {Adolph}}, \bibinfo {author} {\bibfnamefont
  {R.}~\bibnamefont {Andritschke}}, \bibinfo {author} {\bibfnamefont
  {A.}~\bibnamefont {Aquila}}, \bibinfo {author} {\bibfnamefont
  {N.}~\bibnamefont {Berrah}}, \bibinfo {author} {\bibfnamefont
  {C.}~\bibnamefont {Bostedt}}, \bibinfo {author} {\bibfnamefont
  {J.}~\bibnamefont {Bozek}}, \bibinfo {author} {\bibfnamefont
  {N.}~\bibnamefont {Coppola}}, \bibinfo {author} {\bibfnamefont
  {F.}~\bibnamefont {Filsinger}}, \bibinfo {author} {\bibfnamefont
  {H.}~\bibnamefont {Gorke}}, \bibinfo {author} {\bibfnamefont
  {T.}~\bibnamefont {Gorkhover}}, \bibinfo {author} {\bibfnamefont
  {H.}~\bibnamefont {Graafsma}}, \bibinfo {author} {\bibfnamefont
  {L.}~\bibnamefont {Gumprecht}}, \bibinfo {author} {\bibfnamefont
  {A.}~\bibnamefont {Hartmann}}, \bibinfo {author} {\bibfnamefont
  {G.}~\bibnamefont {Hauser}}, \bibinfo {author} {\bibfnamefont
  {S.}~\bibnamefont {Herrmann}}, \bibinfo {author} {\bibfnamefont
  {H.}~\bibnamefont {Hirsemann}}, \bibinfo {author} {\bibfnamefont
  {P.}~\bibnamefont {Holl}}, \bibinfo {author} {\bibfnamefont {A.}~\bibnamefont
  {Hömke}}, \bibinfo {author} {\bibfnamefont {L.}~\bibnamefont {Journel}},
  \bibinfo {author} {\bibfnamefont {C.}~\bibnamefont {Kaiser}}, \bibinfo
  {author} {\bibfnamefont {N.}~\bibnamefont {Kimmel}}, \bibinfo {author}
  {\bibfnamefont {F.}~\bibnamefont {Krasniqi}}, \bibinfo {author}
  {\bibfnamefont {K.-U.}\ \bibnamefont {Kühnel}}, \bibinfo {author}
  {\bibfnamefont {M.}~\bibnamefont {Matysek}}, \bibinfo {author} {\bibfnamefont
  {M.}~\bibnamefont {Messerschmidt}}, \bibinfo {author} {\bibfnamefont
  {D.}~\bibnamefont {Miesner}}, \bibinfo {author} {\bibfnamefont
  {T.}~\bibnamefont {Möller}}, \bibinfo {author} {\bibfnamefont
  {R.}~\bibnamefont {Moshammer}}, \bibinfo {author} {\bibfnamefont
  {K.}~\bibnamefont {Nagaya}}, \bibinfo {author} {\bibfnamefont
  {B.}~\bibnamefont {Nilsson}}, \bibinfo {author} {\bibfnamefont
  {G.}~\bibnamefont {Potdevin}}, \bibinfo {author} {\bibfnamefont
  {D.}~\bibnamefont {Pietschner}}, \bibinfo {author} {\bibfnamefont
  {C.}~\bibnamefont {Reich}}, \bibinfo {author} {\bibfnamefont
  {D.}~\bibnamefont {Rupp}}, \bibinfo {author} {\bibfnamefont {G.}~\bibnamefont
  {Schaller}}, \bibinfo {author} {\bibfnamefont {I.}~\bibnamefont
  {Schlichting}}, \bibinfo {author} {\bibfnamefont {C.}~\bibnamefont
  {Schmidt}}, \bibinfo {author} {\bibfnamefont {F.}~\bibnamefont {Schopper}},
  \bibinfo {author} {\bibfnamefont {S.}~\bibnamefont {Schorb}}, \bibinfo
  {author} {\bibfnamefont {C.-D.}\ \bibnamefont {Schröter}}, \bibinfo {author}
  {\bibfnamefont {J.}~\bibnamefont {Schulz}}, \bibinfo {author} {\bibfnamefont
  {M.}~\bibnamefont {Simon}}, \bibinfo {author} {\bibfnamefont
  {H.}~\bibnamefont {Soltau}}, \bibinfo {author} {\bibfnamefont
  {L.}~\bibnamefont {Strüder}}, \bibinfo {author} {\bibfnamefont
  {K.}~\bibnamefont {Ueda}}, \bibinfo {author} {\bibfnamefont {G.}~\bibnamefont
  {Weidenspointner}}, \bibinfo {author} {\bibfnamefont {R.}~\bibnamefont
  {Santra}}, \bibinfo {author} {\bibfnamefont {J.}~\bibnamefont {Ullrich}},
  \bibinfo {author} {\bibfnamefont {A.}~\bibnamefont {Rudenko}}, \ and\
  \bibinfo {author} {\bibfnamefont {D.}~\bibnamefont {Rolles}},\ }\href
  {\doibase doi:10.1038/nphoton.2012.261} {\bibfield  {journal} {\bibinfo
  {journal} {Nat. Photonics}\ }\textbf {\bibinfo {volume} {6}},\ \bibinfo
  {pages} {858} (\bibinfo {year} {2012})}\BibitemShut {NoStop}%
\bibitem [{\citenamefont {Simonsen}\ \emph {et~al.}(2012)\citenamefont
  {Simonsen}, \citenamefont {S\o{}rng\aa{}rd}, \citenamefont {Nepstad},\ and\
  \citenamefont {F\o{}rre}}]{DAUNTING}%
  \BibitemOpen
  \bibfield  {author} {\bibinfo {author} {\bibfnamefont {A.~S.}\ \bibnamefont
  {Simonsen}}, \bibinfo {author} {\bibfnamefont {S.~A.}\ \bibnamefont
  {S\o{}rng\aa{}rd}}, \bibinfo {author} {\bibfnamefont {R.}~\bibnamefont
  {Nepstad}}, \ and\ \bibinfo {author} {\bibfnamefont {M.}~\bibnamefont
  {F\o{}rre}},\ }\href {\doibase 10.1103/PhysRevA.85.063404} {\bibfield
  {journal} {\bibinfo  {journal} {Phys. Rev. A}\ }\textbf {\bibinfo {volume}
  {85}},\ \bibinfo {pages} {063404} (\bibinfo {year} {2012})}\BibitemShut
  {NoStop}%
\bibitem [{\citenamefont {Malegat}\ \emph {et~al.}(2012)\citenamefont
  {Malegat}, \citenamefont {Bachau}, \citenamefont {Piraux},\ and\
  \citenamefont {Reynal}}]{DAUNTING_1}%
  \BibitemOpen
  \bibfield  {author} {\bibinfo {author} {\bibfnamefont {L.}~\bibnamefont
  {Malegat}}, \bibinfo {author} {\bibfnamefont {H.}~\bibnamefont {Bachau}},
  \bibinfo {author} {\bibfnamefont {B.}~\bibnamefont {Piraux}}, \ and\ \bibinfo
  {author} {\bibfnamefont {F.}~\bibnamefont {Reynal}},\ }\href
  {http://stacks.iop.org/0953-4075/45/i=17/a=175601} {\bibfield  {journal}
  {\bibinfo  {journal} {J. Phys. B: Mol. Opt. Phys.}\ }\textbf {\bibinfo
  {volume} {45}},\ \bibinfo {pages} {175601} (\bibinfo {year}
  {2012})}\BibitemShut {NoStop}%
\bibitem [{\citenamefont {{M. Ilchen and T. Mazza and E. T. Karamatskos and D.
  Markellos and S. Bakhtiarzadeh and A. J. Rafipoor and T. J. Kelly and N.
  Walsh and J. T. Costello and P. O'Keeffe and N. Gerken and M. Martins and P.
  Lambropoulos and M. Meyer}}(2014)}]{NEW_REF}%
  \BibitemOpen
  \bibfield  {author} {\bibinfo {author} {\bibnamefont {{M. Ilchen and T. Mazza
  and E. T. Karamatskos and D. Markellos and S. Bakhtiarzadeh and A. J.
  Rafipoor and T. J. Kelly and N. Walsh and J. T. Costello and P. O'Keeffe and
  N. Gerken and M. Martins and P. Lambropoulos and M. Meyer}}},\ }\href@noop {}
  {} (\bibinfo {year} {2014}),\ \bibinfo {note} {private communication to be
  published}\BibitemShut {NoStop}%
\bibitem [{\citenamefont {{C. A. Ullrich, U. J. Gossmann and E. K. U.
  Gross}}(1995)}]{STRONG}%
  \BibitemOpen
  \bibfield  {author} {\bibinfo {author} {\bibnamefont {{C. A. Ullrich, U. J.
  Gossmann and E. K. U. Gross}}},\ }\href@noop {} {\bibfield  {journal}
  {\bibinfo  {journal} {Ber. Bunsenges. Phys. Chem.}\ }\textbf {\bibinfo
  {volume} {99}},\ \bibinfo {pages} {488} (\bibinfo {year} {1995})}\BibitemShut
  {NoStop}%
\bibitem [{\citenamefont {Ullrich}\ and\ \citenamefont
  {Gross}(1997)}]{STRONG1}%
  \BibitemOpen
  \bibfield  {author} {\bibinfo {author} {\bibfnamefont {C.~A.}\ \bibnamefont
  {Ullrich}}\ and\ \bibinfo {author} {\bibfnamefont {E.~K.~U.}\ \bibnamefont
  {Gross}},\ }\href@noop {} {\bibfield  {journal} {\bibinfo  {journal}
  {Comments At. Mol. Phys.}\ }\textbf {\bibinfo {volume} {33}},\ \bibinfo
  {pages} {211} (\bibinfo {year} {1997})}\BibitemShut {NoStop}%
\bibitem [{\citenamefont {Tong}\ and\ \citenamefont {Chu}(1998)}]{STRONG2}%
  \BibitemOpen
  \bibfield  {author} {\bibinfo {author} {\bibfnamefont {X.-M.}\ \bibnamefont
  {Tong}}\ and\ \bibinfo {author} {\bibfnamefont {S.-I.}\ \bibnamefont {Chu}},\
  }\href@noop {} {\bibfield  {journal} {\bibinfo  {journal} {Phys. Rev. A}\
  }\textbf {\bibinfo {volume} {57}},\ \bibinfo {pages} {452} (\bibinfo {year}
  {1998})}\BibitemShut {NoStop}%
\bibitem [{\citenamefont {Lappas}\ and\ \citenamefont {van
  Leeuwen}(1998)}]{Lappas:1998gc}%
  \BibitemOpen
  \bibfield  {author} {\bibinfo {author} {\bibfnamefont {D.~G.}\ \bibnamefont
  {Lappas}}\ and\ \bibinfo {author} {\bibfnamefont {R.}~\bibnamefont {van
  Leeuwen}},\ }\href@noop {} {\bibfield  {journal} {\bibinfo  {journal} {J.
  Phys. B: At. Mol. Opt. Phys.}\ }\textbf {\bibinfo {volume} {31}},\ \bibinfo
  {pages} {L249} (\bibinfo {year} {1998})}\BibitemShut {NoStop}%
\bibitem [{\citenamefont {Lein}\ \emph {et~al.}(2000)\citenamefont {Lein},
  \citenamefont {Gross},\ and\ \citenamefont {Engel}}]{Lein:2000cu}%
  \BibitemOpen
  \bibfield  {author} {\bibinfo {author} {\bibfnamefont {M.}~\bibnamefont
  {Lein}}, \bibinfo {author} {\bibfnamefont {E.~K.~U.}\ \bibnamefont {Gross}},
  \ and\ \bibinfo {author} {\bibfnamefont {V.}~\bibnamefont {Engel}},\
  }\href@noop {} {\bibfield  {journal} {\bibinfo  {journal} {Phys. Rev. Lett.}\
  }\textbf {\bibinfo {volume} {85}},\ \bibinfo {pages} {4707} (\bibinfo {year}
  {2000})}\BibitemShut {NoStop}%
\bibitem [{\citenamefont {Petersilka}\ and\ \citenamefont
  {Gross}(1999)}]{KNEE}%
  \BibitemOpen
  \bibfield  {author} {\bibinfo {author} {\bibfnamefont {M.}~\bibnamefont
  {Petersilka}}\ and\ \bibinfo {author} {\bibfnamefont {E.~K.~U.}\ \bibnamefont
  {Gross}},\ }\href@noop {} {\bibfield  {journal} {\bibinfo  {journal} {Laser
  Phys.}\ }\textbf {\bibinfo {volume} {9}},\ \bibinfo {pages} {105} (\bibinfo
  {year} {1999})}\BibitemShut {NoStop}%
\bibitem [{\citenamefont {Bauer}\ and\ \citenamefont
  {Ceccherini}(2001)}]{KNEE_AGAIN}%
  \BibitemOpen
  \bibfield  {author} {\bibinfo {author} {\bibfnamefont {D.}~\bibnamefont
  {Bauer}}\ and\ \bibinfo {author} {\bibfnamefont {F.}~\bibnamefont
  {Ceccherini}},\ }\href {\doibase 10.1364/OE.8.000377} {\bibfield  {journal}
  {\bibinfo  {journal} {Opt. Express}\ }\textbf {\bibinfo {volume} {8}},\
  \bibinfo {pages} {377} (\bibinfo {year} {2001})}\BibitemShut {NoStop}%
\bibitem [{\citenamefont {Runge}\ and\ \citenamefont
  {Gross}(1984)}]{RUNGEGROSS}%
  \BibitemOpen
  \bibfield  {author} {\bibinfo {author} {\bibfnamefont {E.}~\bibnamefont
  {Runge}}\ and\ \bibinfo {author} {\bibfnamefont {E.~K.~U.}\ \bibnamefont
  {Gross}},\ }\href@noop {} {\bibfield  {journal} {\bibinfo  {journal} {Phys.
  Rev. Lett}\ }\textbf {\bibinfo {volume} {52}},\ \bibinfo {pages} {997}
  (\bibinfo {year} {1984})}\BibitemShut {NoStop}%
\bibitem [{\citenamefont {Kohn}\ and\ \citenamefont {Sham}(1965)}]{KS}%
  \BibitemOpen
  \bibfield  {author} {\bibinfo {author} {\bibfnamefont {W.}~\bibnamefont
  {Kohn}}\ and\ \bibinfo {author} {\bibfnamefont {L.~J.}\ \bibnamefont
  {Sham}},\ }\href@noop {} {\bibfield  {journal} {\bibinfo  {journal} {Phys.
  Rev.}\ }\textbf {\bibinfo {volume} {140}},\ \bibinfo {pages} {A1133}
  (\bibinfo {year} {1965})}\BibitemShut {NoStop}%
\bibitem [{\citenamefont {Perdew}\ and\ \citenamefont {Wang}(1992)}]{LDA1}%
  \BibitemOpen
  \bibfield  {author} {\bibinfo {author} {\bibfnamefont {J.~P.}\ \bibnamefont
  {Perdew}}\ and\ \bibinfo {author} {\bibfnamefont {Y.}~\bibnamefont {Wang}},\
  }\href@noop {} {\bibfield  {journal} {\bibinfo  {journal} {Phys. Rev. B}\
  }\textbf {\bibinfo {volume} {45}},\ \bibinfo {pages} {13244} (\bibinfo {year}
  {1992})}\BibitemShut {NoStop}%
\bibitem [{\citenamefont {{J. P. Perdew, K. Burke and M.
  Ernzerhof}}(1996)}]{PBE}%
  \BibitemOpen
  \bibfield  {author} {\bibinfo {author} {\bibnamefont {{J. P. Perdew, K. Burke
  and M. Ernzerhof}}},\ }\href@noop {} {\bibfield  {journal} {\bibinfo
  {journal} {Phys. Rev. Lett.}\ }\textbf {\bibinfo {volume} {77}},\ \bibinfo
  {pages} {3865} (\bibinfo {year} {1996})}\BibitemShut {NoStop}%
\bibitem [{\citenamefont {van Leeuwen}\ and\ \citenamefont
  {Baerends}(1994)}]{LB94}%
  \BibitemOpen
  \bibfield  {author} {\bibinfo {author} {\bibfnamefont {R.}~\bibnamefont {van
  Leeuwen}}\ and\ \bibinfo {author} {\bibfnamefont {E.~J.}\ \bibnamefont
  {Baerends}},\ }\href@noop {} {\bibfield  {journal} {\bibinfo  {journal}
  {Phys. Rev. A}\ }\textbf {\bibinfo {volume} {49}},\ \bibinfo {pages} {2421}
  (\bibinfo {year} {1994})}\BibitemShut {NoStop}%
\bibitem [{\citenamefont {Andrade}\ and\ \citenamefont
  {Aspuru-Guzik}(2011)}]{XCLDA}%
  \BibitemOpen
  \bibfield  {author} {\bibinfo {author} {\bibfnamefont {X.}~\bibnamefont
  {Andrade}}\ and\ \bibinfo {author} {\bibfnamefont {A.}~\bibnamefont
  {Aspuru-Guzik}},\ }\href@noop {} {\bibfield  {journal} {\bibinfo  {journal}
  {Phys. Rev. Lett.}\ }\textbf {\bibinfo {volume} {107}},\ \bibinfo {pages}
  {183002} (\bibinfo {year} {2011})}\BibitemShut {NoStop}%
\bibitem [{\citenamefont {Ullrich}(2000)}]{Ullrich:2000tm}%
  \BibitemOpen
  \bibfield  {author} {\bibinfo {author} {\bibfnamefont {C.~A.}\ \bibnamefont
  {Ullrich}},\ }\href@noop {} {\bibfield  {journal} {\bibinfo  {journal} {J.
  Mol. Struct. (THEOCHEM)}\ }\textbf {\bibinfo {volume} {501-502}},\ \bibinfo
  {pages} {315} (\bibinfo {year} {2000})}\BibitemShut {NoStop}%
\bibitem [{\citenamefont {{M. A. L. Marques, A. Castro, G.F. Bertsch and A.
  Rubio}}(2003)}]{OCTOPUS1}%
  \BibitemOpen
  \bibfield  {author} {\bibinfo {author} {\bibnamefont {{M. A. L. Marques, A.
  Castro, G.F. Bertsch and A. Rubio}}},\ }\href@noop {} {\bibfield  {journal}
  {\bibinfo  {journal} {Comput. Phys. Commun.}\ }\textbf {\bibinfo {volume}
  {151}},\ \bibinfo {pages} {60} (\bibinfo {year} {2003})}\BibitemShut
  {NoStop}%
\bibitem [{\citenamefont {{A. Castro and M.A.L. Marques and H. Appel and M.
  Oliveira and C.A. Rozzi and X. Andrade and F. Lorenzen and E.K.U Gross and A.
  Rubio}}(2006)}]{OCTOPUS2}%
  \BibitemOpen
  \bibfield  {author} {\bibinfo {author} {\bibnamefont {{A. Castro and M.A.L.
  Marques and H. Appel and M. Oliveira and C.A. Rozzi and X. Andrade and F.
  Lorenzen and E.K.U Gross and A. Rubio}}},\ }\href@noop {} {\bibfield
  {journal} {\bibinfo  {journal} {Phys. Stat. Sol. B}\ }\textbf {\bibinfo
  {volume} {243}},\ \bibinfo {pages} {2465} (\bibinfo {year}
  {2006})}\BibitemShut {NoStop}%
\bibitem [{\citenamefont {Andrade}\ \emph {et~al.}(2012)\citenamefont
  {Andrade}, \citenamefont {Alberdi-Rodriguez}, \citenamefont {Strubbe},
  \citenamefont {Oliveira}, \citenamefont {Nogueira}, \citenamefont {Castro},
  \citenamefont {Muguerza}, \citenamefont {Arruabarrena}, \citenamefont
  {Louie}, \citenamefont {Aspuru-Guzik}, \citenamefont {Rubio},\ and\
  \citenamefont {Marques}}]{OCTOPUS3}%
  \BibitemOpen
  \bibfield  {author} {\bibinfo {author} {\bibfnamefont {X.}~\bibnamefont
  {Andrade}}, \bibinfo {author} {\bibfnamefont {J.}~\bibnamefont
  {Alberdi-Rodriguez}}, \bibinfo {author} {\bibfnamefont {D.~A.}\ \bibnamefont
  {Strubbe}}, \bibinfo {author} {\bibfnamefont {M.~J.~T.}\ \bibnamefont
  {Oliveira}}, \bibinfo {author} {\bibfnamefont {F.}~\bibnamefont {Nogueira}},
  \bibinfo {author} {\bibfnamefont {A.}~\bibnamefont {Castro}}, \bibinfo
  {author} {\bibfnamefont {J.}~\bibnamefont {Muguerza}}, \bibinfo {author}
  {\bibfnamefont {A.}~\bibnamefont {Arruabarrena}}, \bibinfo {author}
  {\bibfnamefont {S.~G.}\ \bibnamefont {Louie}}, \bibinfo {author}
  {\bibfnamefont {A.}~\bibnamefont {Aspuru-Guzik}}, \bibinfo {author}
  {\bibfnamefont {A.}~\bibnamefont {Rubio}}, \ and\ \bibinfo {author}
  {\bibfnamefont {M.~A.~L.}\ \bibnamefont {Marques}},\ }\href
  {http://stacks.iop.org/0953-8984/24/i=23/a=233202} {\bibfield  {journal}
  {\bibinfo  {journal} {Journal of Physics: Condensed Matter}\ }\textbf
  {\bibinfo {volume} {24}},\ \bibinfo {pages} {233202} (\bibinfo {year}
  {2012})}\BibitemShut {NoStop}%
\bibitem [{\citenamefont {{U. De Giovannini, A. H. Larsen and
  A.Rubio}}(2014)}]{PUBLISH}%
  \BibitemOpen
  \bibfield  {author} {\bibinfo {author} {\bibnamefont {{U. De Giovannini, A.
  H. Larsen and A.Rubio}}},\ }\href@noop {} {\enquote {\bibinfo {title}
  {Modeling electron dynamics coupled to continuum states in finite volumes},}\
  } (\bibinfo {year} {2014}),\ \bibinfo {note} {unpublished}\BibitemShut
  {NoStop}%
\bibitem [{\citenamefont {Troullier}\ and\ \citenamefont {Martins}(1991)}]{TM}%
  \BibitemOpen
  \bibfield  {author} {\bibinfo {author} {\bibfnamefont {N.}~\bibnamefont
  {Troullier}}\ and\ \bibinfo {author} {\bibfnamefont {J.~L.}\ \bibnamefont
  {Martins}},\ }\href@noop {} {\bibfield  {journal} {\bibinfo  {journal} {Phys.
  Rev. B}\ }\textbf {\bibinfo {volume} {43}},\ \bibinfo {pages} {1993}
  (\bibinfo {year} {1991})}\BibitemShut {NoStop}%
\bibitem [{\citenamefont {Oliveira}\ and\ \citenamefont
  {Nogueira}(2008)}]{APE}%
  \BibitemOpen
  \bibfield  {author} {\bibinfo {author} {\bibfnamefont {M.}~\bibnamefont
  {Oliveira}}\ and\ \bibinfo {author} {\bibfnamefont {F.}~\bibnamefont
  {Nogueira}},\ }\href@noop {} {\bibfield  {journal} {\bibinfo  {journal}
  {Comput. Phys. Commun.}\ }\textbf {\bibinfo {volume} {178}},\ \bibinfo
  {pages} {524} (\bibinfo {year} {2008})}\BibitemShut {NoStop}%
\bibitem [{\citenamefont {Fuks}\ \emph {et~al.}(2011)\citenamefont {Fuks},
  \citenamefont {Helbig}, \citenamefont {Tokatly},\ and\ \citenamefont
  {Rubio}}]{Fuks:2011hp}%
  \BibitemOpen
  \bibfield  {author} {\bibinfo {author} {\bibfnamefont {J.~I.}\ \bibnamefont
  {Fuks}}, \bibinfo {author} {\bibfnamefont {N.}~\bibnamefont {Helbig}},
  \bibinfo {author} {\bibfnamefont {I.~V.}\ \bibnamefont {Tokatly}}, \ and\
  \bibinfo {author} {\bibfnamefont {A.}~\bibnamefont {Rubio}},\ }\href@noop {}
  {\bibfield  {journal} {\bibinfo  {journal} {Phys. Rev. B}\ }\textbf {\bibinfo
  {volume} {84}},\ \bibinfo {pages} {075107} (\bibinfo {year}
  {2011})}\BibitemShut {NoStop}%
\bibitem [{\citenamefont {Fuks}\ \emph {et~al.}(2013)\citenamefont {Fuks},
  \citenamefont {Elliott}, \citenamefont {Rubio},\ and\ \citenamefont
  {Maitra}}]{Fuks:2013fn}%
  \BibitemOpen
  \bibfield  {author} {\bibinfo {author} {\bibfnamefont {J.~I.}\ \bibnamefont
  {Fuks}}, \bibinfo {author} {\bibfnamefont {P.}~\bibnamefont {Elliott}},
  \bibinfo {author} {\bibfnamefont {A.}~\bibnamefont {Rubio}}, \ and\ \bibinfo
  {author} {\bibfnamefont {N.~T.}\ \bibnamefont {Maitra}},\ }\href@noop {}
  {\bibfield  {journal} {\bibinfo  {journal} {J. Phys. Chem. Lett.}\ }\textbf
  {\bibinfo {volume} {4}},\ \bibinfo {pages} {735} (\bibinfo {year}
  {2013})}\BibitemShut {NoStop}%
\bibitem [{\citenamefont {Elliott}\ \emph {et~al.}(2012)\citenamefont
  {Elliott}, \citenamefont {Fuks}, \citenamefont {Rubio},\ and\ \citenamefont
  {Maitra}}]{PRL_ELLIOT}%
  \BibitemOpen
  \bibfield  {author} {\bibinfo {author} {\bibfnamefont {P.}~\bibnamefont
  {Elliott}}, \bibinfo {author} {\bibfnamefont {J.~I.}\ \bibnamefont {Fuks}},
  \bibinfo {author} {\bibfnamefont {A.}~\bibnamefont {Rubio}}, \ and\ \bibinfo
  {author} {\bibfnamefont {N.~T.}\ \bibnamefont {Maitra}},\ }\href@noop {}
  {\bibfield  {journal} {\bibinfo  {journal} {Phys. Rev. Lett.}\ }\textbf
  {\bibinfo {volume} {109}},\ \bibinfo {pages} {266404} (\bibinfo {year}
  {2012})}\BibitemShut {NoStop}%
\bibitem [{\citenamefont {Tong}\ and\ \citenamefont {Chu}(1997)}]{CHU}%
  \BibitemOpen
  \bibfield  {author} {\bibinfo {author} {\bibfnamefont {X.-M.}\ \bibnamefont
  {Tong}}\ and\ \bibinfo {author} {\bibfnamefont {S.-I.}\ \bibnamefont {Chu}},\
  }\href {\doibase 10.1103/PhysRevA.55.3406} {\bibfield  {journal} {\bibinfo
  {journal} {Phys. Rev. A}\ }\textbf {\bibinfo {volume} {55}},\ \bibinfo
  {pages} {3406} (\bibinfo {year} {1997})}\BibitemShut {NoStop}%
\bibitem [{\citenamefont {Leforestier}\ and\ \citenamefont
  {Wyatt}(1983)}]{Leforestier:1983gka}%
  \BibitemOpen
  \bibfield  {author} {\bibinfo {author} {\bibfnamefont {C.}~\bibnamefont
  {Leforestier}}\ and\ \bibinfo {author} {\bibfnamefont {R.~E.}\ \bibnamefont
  {Wyatt}},\ }\href@noop {} {\bibfield  {journal} {\bibinfo  {journal} {J.
  Chem. Phys.}\ }\textbf {\bibinfo {volume} {78}},\ \bibinfo {pages} {2334}
  (\bibinfo {year} {1983})}\BibitemShut {NoStop}%
\bibitem [{\citenamefont {Kosloff}\ and\ \citenamefont
  {Kosloff}(1986)}]{Kosloff:1986eta}%
  \BibitemOpen
  \bibfield  {author} {\bibinfo {author} {\bibfnamefont {R.}~\bibnamefont
  {Kosloff}}\ and\ \bibinfo {author} {\bibfnamefont {D.}~\bibnamefont
  {Kosloff}},\ }\href@noop {} {\bibfield  {journal} {\bibinfo  {journal} {J.
  Comput. Phys.}\ }\textbf {\bibinfo {volume} {63}},\ \bibinfo {pages} {363}
  (\bibinfo {year} {1986})}\BibitemShut {NoStop}%
\bibitem [{\citenamefont {Samson}\ and\ \citenamefont
  {Stolte}(2002)}]{Samson:2002hx}%
  \BibitemOpen
  \bibfield  {author} {\bibinfo {author} {\bibfnamefont {J.~A.~R.}\
  \bibnamefont {Samson}}\ and\ \bibinfo {author} {\bibfnamefont {W.~C.}\
  \bibnamefont {Stolte}},\ }\href@noop {} {\bibfield  {journal} {\bibinfo
  {journal} {J. Electron Spectrosc.}\ }\textbf {\bibinfo {volume} {123}},\
  \bibinfo {pages} {265} (\bibinfo {year} {2002})}\BibitemShut {NoStop}%
\bibitem [{\citenamefont {Marques}\ \emph {et~al.}(2011)\citenamefont
  {Marques}, \citenamefont {Maitra}, \citenamefont {Nogueira}, \citenamefont
  {Gross},\ and\ \citenamefont {Rubio}}]{Marques:2011ud}%
  \BibitemOpen
  \bibfield  {author} {\bibinfo {author} {\bibfnamefont {M.~A.~L.}\
  \bibnamefont {Marques}}, \bibinfo {author} {\bibfnamefont {N.~T.}\
  \bibnamefont {Maitra}}, \bibinfo {author} {\bibfnamefont {F.}~\bibnamefont
  {Nogueira}}, \bibinfo {author} {\bibfnamefont {E.~K.~U.}\ \bibnamefont
  {Gross}}, \ and\ \bibinfo {author} {\bibfnamefont {A.}~\bibnamefont
  {Rubio}},\ }\href@noop {} {\emph {\bibinfo {title} {{Fundamentals of
  Time-Dependent Density Functional Theory, Lecture Notes in Physics, Vol.
  837}}}}\ (\bibinfo  {publisher} {Springer},\ \bibinfo {year}
  {2011})\BibitemShut {NoStop}%
\end{thebibliography}

%

\end{document}